\documentclass[preprint]{aastex63}

\usepackage{xspace}
\usepackage{graphicx}
\usepackage{natbib}
\usepackage{url}
\usepackage{bm}
\usepackage{amsmath}
\usepackage{balance}
\usepackage[maxfloats=256]{morefloats}

\newcommand\urltilda{\kern -.15em\lower .7ex\hbox{\~{}}\kern .04em}

\newcommand{\adp}{\ensuremath{\rho_{\text{AD}}(x,y)}}
\newcommand{\adpb}{\ensuremath{\rho_{\text{AD}}\bigl({\tilde x}^{(i)},{\tilde y}^{(i)}\bigr)}}

\newcommand{\adpba}{\ensuremath{\rho_{\text{AD}}^{\text{boot}}}}

\begin{document}

\title{Gender Systematics in the NRAO Proposal Review System}

\author[0000-0003-4320-6378]{Gareth Hunt}
\affiliation{National Radio Astronomy Observatory, 520 Edgemont Rd.,
  Charlottesville, VA 22903, USA}

\author{Frederic R. Schwab}
\affiliation{National Radio Astronomy Observatory, 520 Edgemont Rd.,
  Charlottesville, VA 22903, USA}

\author{P. A. Henning}
\affiliation{National Radio Astronomy
  Observatory, P.O. Box O, Socorro, NM 87801, USA}
\affiliation{Department of Physics and Astronomy, MSC07 4220, 1
  University of New Mexico, Albuquerque NM 87131, USA}

\author[0000-0002-2465-7803]{Dana S. Balser}
\affiliation{National Radio Astronomy Observatory, 520 Edgemont Rd.,
  Charlottesville, VA 22903, USA}

\begin{abstract}

  Several recent investigations indicate the existence of
  gender-related systematic trends in the peer review of proposals for
  observations on astronomical facilities.  This includes the National
  Radio Astronomy Observatory (NRAO) where there is evidence of a
  gender imbalance in the rank of proposals with male principal
  investigators (PIs) favored over female PIs.  Since semester 2017A
  (17A), the NRAO has taken the following steps: (1) inform science
  review panels (SRPs) and the telescope time allocation committee
  (TAC) about the gender imbalance; and (2) increase the female
  representation on SRPs and the TAC to reflect the community
  demographics.  Here we analyze SRP normalized rank-ordered scores,
  or linear ranks, by PI gender for NRAO observing proposals from
  semesters 12A--21A\@.  We use bootstrap resampling to generate
  modeled distributions and the Anderson-Darling (AD) test to evaluate
  the probability that the linear rank distributions for male and
  female PIs are drawn from the same parent sample.  We find that
  between semesters 12A--17A that male PIs are favored over female PIs
  (AD {\it p}-value 0.0084), whereas between semesters 17B--21A female
  PIs are favored over male PIs, but at a lower significance (AD {\it
    p}-value 0.11).  Therefore the gender imbalance is currently being
  ameliorated, but this imbalance may have been reversed.  Regardless,
  we plan to adopt a dual-anonymous approach to proposal review to
  reduce the possibility of bias to occur.
    
\end{abstract}


\section{Introduction}\label{sec:intro}

Access to a national facility such as the NRAO is often critical to
the success of a scientist's research program.  The NRAO therefore
allocates telescope time through a competitive peer review process as
do many other observatories.  Studies show that unconscious bias can
play a significant role when evaluating proposals, but the situation
is complex and other factors can explain systematic trends in proposal
success rates \citep[e.g., see][and references within]{johnson20}.  A
broad representation of our user community should produce the most
compelling science, and therefore identifying systematic trends in the
proposal review process is important.

\citet{reid14} found that male\footnote{We are aware that there has
  been some discussion on use of ``male/female'' instead of
  ``men/women.''  Here we follow previous studies of gender
  systematics and use ``male/female.''}  PIs had a higher success rate
than female PIs for Hubble Space Telescope (HST) observing proposals
from cycles 11--21.  The gender imbalance in any single cycle was
subtle and not immediately obvious, but a pattern was noticeable when
several successive proposal cycles were combined.  These differences
do not depend on the geographic location of the PI nor on the gender
distribution of the reviewers, but review panels with a higher average
seniority produced the lower success rates for proposals with female
PIs.  This led the Space Telescope Science Institute (STScI) to adopt
a dual-anonymous proposal review process starting in cycle 26.
\citet{johnson20} investigated reviewer bias and found that female PIs
were ranked lower than male PIs by male reviewers before, but not
after, the dual-anonymous process was adopted.  Additional data are
needed to confirm this trend.

The findings by \citet{reid14} prompted studies at other
observatories.  \citet{patat16} analyzed gender systematics in
telescope time allocation at the European Southern Observatory (ESO)
over a period of eight years.  Similar to HST, \citet{patat16} found
that male PIs had a higher success rate than female PIs, but this
difference could mostly be attributed to seniority.  For example, only
34\% of female PIs are professional astronomers, which do not include
post-docs or students, compared with 53\% for male PIs.  Nevertheless,
after accounting for seniority there remained a small, but
significant, gender imbalance.

\citet{lonsdale16} studied gender systematics in the facilities
operated by Associated Universities Inc.\ (AUI).  These include the
Atacama Large Millimeter/submillimeter Array (ALMA)\footnote{ALMA is
  operated by the NRAO/AUI in partnership with ESO and the National
  Astronomical Observatories of Japan (NAOJ), in cooperation with the
  Republic of Chile.} and AUI's North American (NA) facilities: the
Very Large Array (VLA), the Very Long Baseline Array (VLBA), and the
Green Bank Telescope (GBT)\footnote{Since semester 17A the GBT has
  been operated by the Green Bank Observatory (GBO)/AUI.}.  ALMA
cycles 2--4 and NRAO semesters 12A--17A were analyzed.
\citet{lonsdale16} found a significant gender imbalance where male PIs
were favored over female PIs.  For ALMA the rank distributions between
male and female PIs were not the same with a high reliability.  A
similar result was found for AUI's NA facilities but with less
significance.  The was no correlation between gender rankings and the
gender distribution on the review panels.  The data to study the
seniority of PIs was not available.

\citet{carpenter20} expanded the gender systematics study of ALMA to
cycles 0--6.  The analysis considered both review stage 1, which
includes preliminary scores, and review stage 2, which are final
scores after a face-to-face discussion within each panel.  Regardless
of regional demographics, male PIs were found to have a higher
proposal success rate than female PIs for each cycle.  In any given
cycle, however, the result is not significant ($> 3\sigma$).
Comparison of stage 1 and stage 2 results revealed no significant
changes in the distribution of rank by gender, region, or seniority.
He concluded that any systematics in the proposal ranking is not
introduced from the face-to-face discussions.

\citet{hunt19, hunt21} extended the analysis of \citet{lonsdale16} for
AUI's NA facilities to semesters 12A--21A\@.  Using the same procedures
they initially found similar trends, that male PIs were favored over
female PIs, but in later semesters this gender imbalance appeared to
be ameliorated.  During semesters 17B--21A there appears to be a
slight imbalance that favors female PIs over male PIs.  Here we
perform a more rigorous analysis of the data for AUI's NA facilities
over semesters 12A--21A to help understand the significance of these
results.

\section{The NRAO Proposal Review System}\label{sec:review}

The NRAO runs the proposal process for AUI's NA facilities which
includes the VLA, VLBA, and GBT \citep[see][]{schwab15}.  Observing
proposals are submitted twice per year with nominal deadlines of 1
August (semester A), for observations to be made the following year
from February to July, and 1 February (semester B), for observations
to be made from August to the following January.  Proposals are
evaluated based on their scientific merit and technical feasibility
using a panel-based review system.  There are currently nine science
review panels (SRPs) which are divided into different areas of
science.  Each SRP consists of at least 5 reviewers plus a chair.
Proposals are initially reviewed and assigned an individual raw score
from 0.1--9.9 (lower scores are better).  Once the individual reviews
are complete, the scores are normalized for each reviewer to have a
mean of 5.0 and a standard deviation of 2.0.  Each SRP will then meet
remotely to discuss the proposals and form a consensus review.  At
this stage a technical review of the feasibility of the proposed
observations, performed by observatory staff, is made available to the
SRP.  During the SRP meeting, scores may be adjusted based on the
discussion.  After the consensus reviews are completed, the proposals
in each SRP are rank-ordered and normalized to obtain what we call a
linear rank. The linear rank for the {\it i}$^{\rm th}$ proposal is
$10\,r_{\it i}/n$, where $r_{\it i}$ is the rank associated with that
proposal and $n$ is the number of proposals reviewed by the SRP.

The TAC consists of the SRP chairs, with one serving as the TAC chair,
and meets face-to-face after the SRP meetings.\footnote{During
  semesters 20B--22A the TAC met remotely because of Covid-19; in
  future semesters the TAC meeting will be hybrid.}  The TAC considers
the linear ranks, together with any technical, resource, or scheduling
constraints to assign a scheduling priority.  The TAC does not
reevaluate the scientific merit of the proposal.  Priority A is the
highest scheduling priority and the observations will almost certainly
be scheduled.  Priority B is the next highest scheduling priority and
the observations will be scheduled on a best effort basis.  Priority C
is the lowest scheduling priority (e.g., filler time).  Priority N
will not be scheduled.  The TAC does not allocate telescope time but
makes recommendations to the observatory Directors.  A Directors'
review is held each semester to evaluate the TAC's recommendations.  A
final program is then approved.  Therefore, a successful proposal
might be defined as a proposal that receives a scheduling priority A
or B\@.  But this depends on resource and scheduling constraints and
therefore here we use the linear rank as the measure of success.

\begin{figure}
  \centering
  \includegraphics[angle=0,scale=0.80]{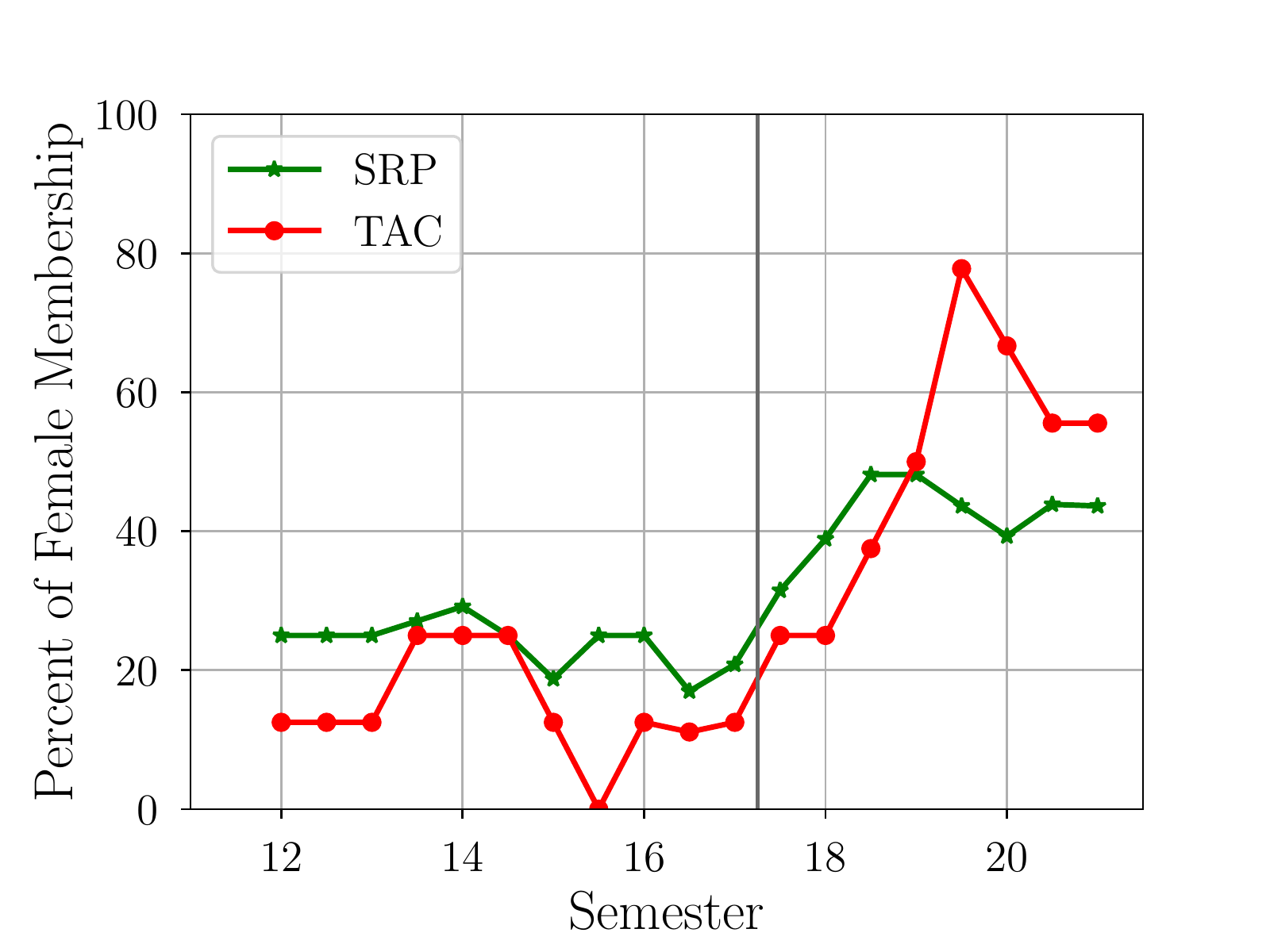}
  \caption{The gender distribution of SRPs and the TAC as a function
    of semester.  The gray vertical line indicates when the NRAO began
    (1) informing the SRPs and TAC of the gender bias; and (2)
    increasing female members on these panels to reflect the community
    demographics.}
\label{fig:gender}
\end{figure}

In response to the findings in \citet{lonsdale16} two significant
administrative changes were made to the review process starting with
semester 17B\@. (1) At the beginning of each SRP and TAC meeting the
status of the gender imbalance is discussed and the reviewers are
reminded that rankings and decisions must reflect only scientific
merit, technical feasibility, and operational constraints. (2) The
NRAO set goals to populate the SRPs and TAC with a gender distribution
that reflects the community demographics ($\sim\,$30--35\% female
membership for NRAO users) and furthermore to arrange that there
should be at least 2 female reviewers on each SRP and the
TAC. Figure~\ref{fig:gender} shows the female fraction of SRPs and TAC
by semester.  These goals were achieved after only a few semesters.

\clearpage

\section{Data and Analysis}\label{sec:data}

The data consist of the linear ranks and the PI gender for each
proposal.\footnote{See go.nrao.edu/prop-gender for access to the
  data.}  We also track the gender distribution of the SRPs and TAC.
The PI gender is taken from the user account profile.  Historically
the user was able to specify their gender as ``male'' or ``female.''
More recently we have allowed the user to specify their gender as
``male,'' ``female,'' or ``self-identify.''  Regardless, the selection
of gender is optional since we are not permitted to request the gender
of any person using our instruments.  Only $\sim\,$65\% of users
specify their gender.  Most other PI gender assignments were
determined by performing literature or web searches; a few were
established by asking colleagues who had been
collaborators.\footnote{We are sensitive to the fact that some PIs
  will never want their gender to be identified.  In the future, we
  will encourage users to specify their gender to provide us with a
  statistically significant sample using only the self-declarations.}
Adding these genders to the database increases the percentage of PIs
with a defined gender to 99.6\%.  A few genders are thus obviously
still not determined, and the corresponding proposals are not included
in our analysis. This can also lead to a slight inconsistency with
time, since registrants may choose to add their gender to their
profiles and others' genders may be determined later.

We divide the data into 32 different data subsets for analysis. We
consider each semester (19 subsets); semesters 12A--17A and 17B--21A
(2 subsets); each year (10 subsets); and all years (1 subset).  For
each data subset we calculate both the cumulative distribution
function (CDF) and the 25$^{\rm th}$, 50$^{\rm th}$, and 75$^{\rm th}$
percentiles for the male and female PI linear ranks.  These
percentiles are commonly referred to as the ``quartile values'' (i.e.,
Q1, Q2, and Q3).  To assess the uncertainty in the linear rank
distributions, we apply bootstrap resampling \citep{efron82}.  To do
this we generate 10,000 simulated distributions of the male and female
PI linear ranks.  Each distribution consists of $N$ data points which
are randomly drawn from the original data with replacement, where $N$
is the number of linear ranks in the sample.  So the simulated
distributions will miss some linear rank values from the original data
and have some duplicates, triplicates, etc.

We use the Anderson-Darling (AD) test to evaluate the probability that
the observed linear rank distributions for male and female PIs are
drawn from the same parent sample.  The AD statistic is considered to
be more reliable than the Kolmogorov-Smirnov or Cram\'{e}r-von Mises
tests \citep[cf.,][]{babu06}.  The AD {\it p}-value is denoted as
\adp, where $x$ is the list of female proposers' linear ranks and $y$
the list of the male proposers'.  To assess the uncertainty in the AD
test results we again apply bootstrap resampling where
${\tilde x}^{(i)}$ represents the $i^{\text{th}}$ of $m = 10,000$
bootstrap resampling of the $x$ values, and similarly
${\tilde y}^{(i)}$ represents the $i^{\text{th}}$ resampling of the
$y$ values.  Then \adpb\ is the AD {\it p}-value of the
$i^{\text{th}}$ resample, and
$\rho_{\text{AD}}^{\text{boot}}\equiv {1\over m}\sum_{i=1}^m
\rho_{\text{AD}}(\tilde x^{(i)},\tilde y^{(i)})$ is the mean of the
bootstrap AD {\it p}-values.

A low \adp\ indicates that the two distributions are not from a common
parent distribution, whereas a higher \adp\ suggests there is a
degree of commonality.  To broadly determine if there are gender
imbalances in the NRAO proposal system, we assign an imbalance key (IK)
indicating which gender, if any, is favored using the following
criteria:

\begin{itemize}

\item Upper Case Letter (M or F): if $\rho_{\text{AD}}(x,y) \le 0.1$ (very
  suggestive).

\item Lower Case Letter (m or f): if $\rho_{\text{AD}}(x,y) \le 0.2$ (suggestive).

\item Blank: if $\rho_{\text{AD}}(x,y) > 0.2$.

\end{itemize}

When there is an imbalance the \adp\ does not indicate which
distribution has the lower (better) linear ranks. This is determined
by inspecting the male and female PI linear rank distributions.

\section{Results}\label{sec:results}

\begin{figure}
  \centering
  \includegraphics[angle=0,scale=0.9]{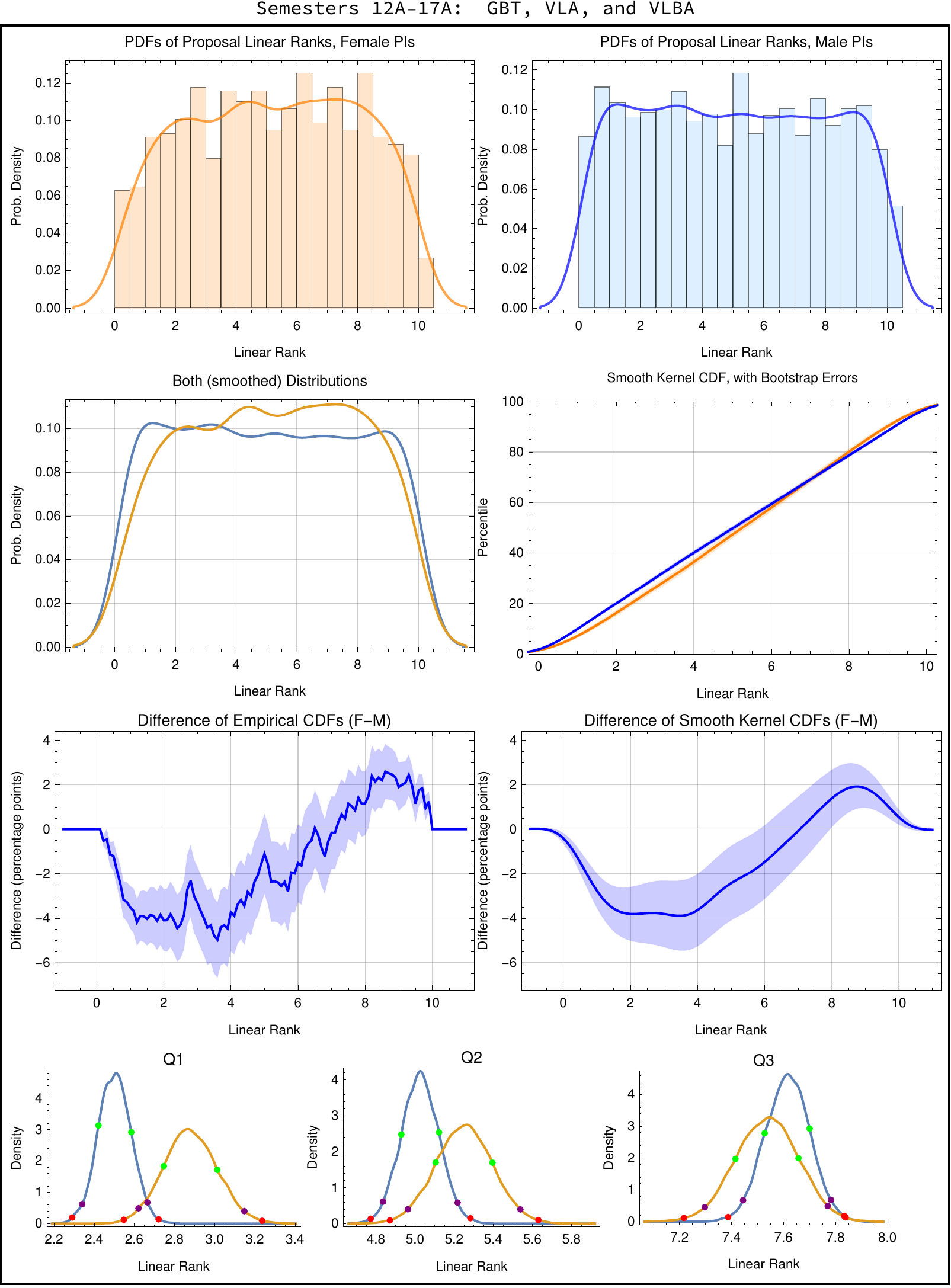}
  \caption{Statistics for semesters 12A--17A\@.  {\it Top:} histogram of
    the linear rank for female PIs (left, orange) and male PIs (right,
    blue).  The curves are the smooth kernel histograms. {\it
      Middle-top:} both male and female PI smooth kernel histograms on
    the same plot (left), and smooth kernel CDFs (right).  The shaded
    regions are the uncertainties from bootstrap resampling.  {\it
      Middle-bottom:} the difference (female minus male) of the
    empirical CDFs (left) and the smooth kernel CDFs (right). The
    shaded regions are the uncertainties from bootstrap resampling.
    {\it Bottom:} the quartile distributions from bootstrap resampling
    of female PI (orange) and male PI (blue) linear ranks.  Plotted
    are the 25$^{\rm th}$ (Q1), 50$^{\rm th}$ (Q2), and 75$^{\rm th}$
    (Q3) percentiles.  The dots are the 68\% (green), 95\% (purple),
    and 99\% (red) confidence levels of these distributions.}
\label{fig:stats_12a-17a}
\end{figure}

\begin{figure}
  \centering
  \includegraphics[angle=0,scale=0.9]{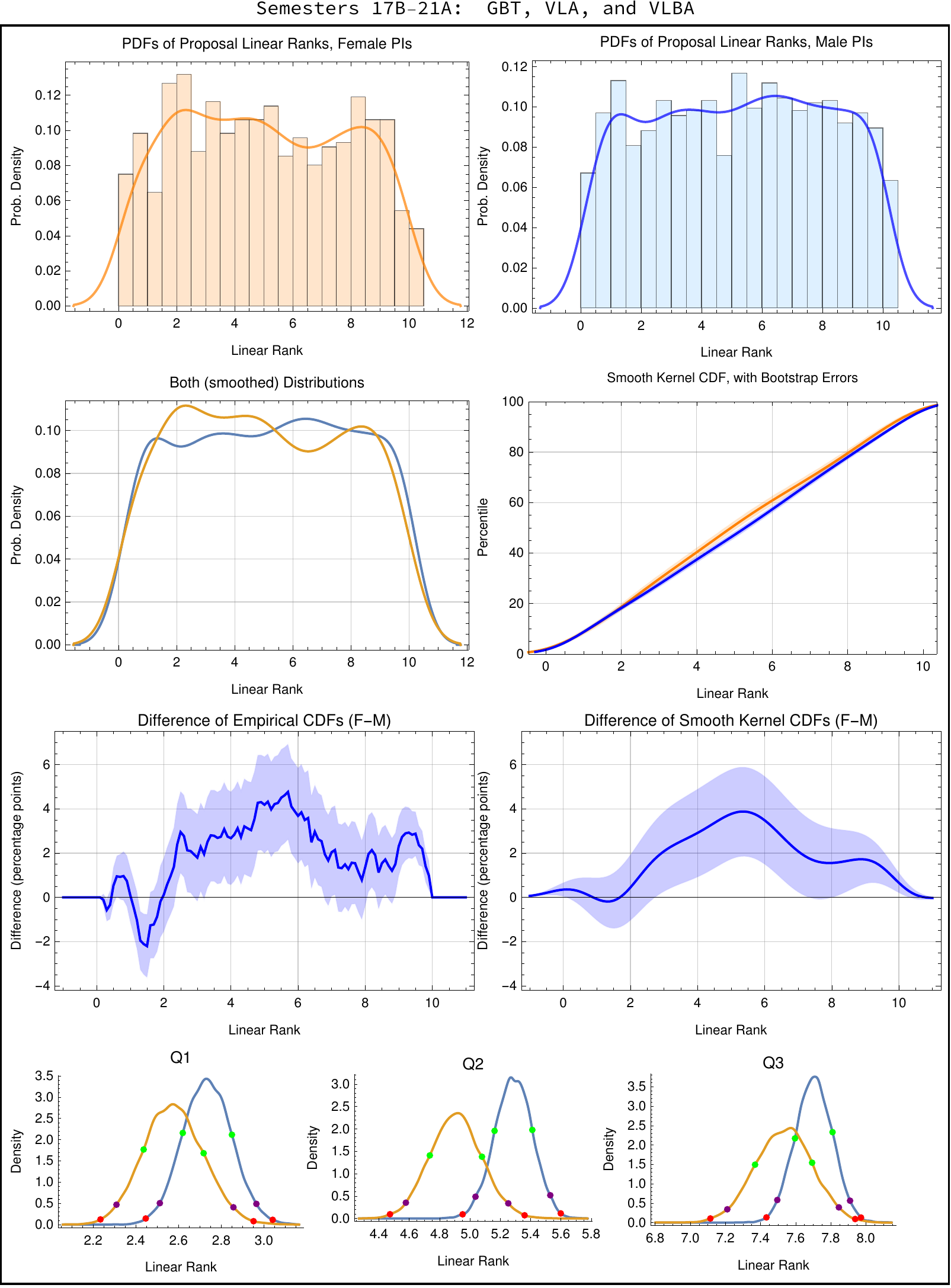}
  \caption{Statistics for semesters 17B--21A\@. See
    Figure~\ref{fig:stats_12a-17a} for details.}
\label{fig:stats_17b-21a}
\end{figure}

We generate a series of plots for each data subset that includes
histograms, CDFs, and quartile distributions of the male and female PI
linear ranks.  Figures~\ref{fig:stats_12a-17a} and
\ref{fig:stats_17b-21a} show the results for semesters 12A--17A and
17B--21A, respectively.  The smoothed distributions are produced using
kernel density estimation \citep[KDE,][]{silverman96}.\footnote{See
  https://en.wikipedia.org/wiki/Kernel\_density\_estimation for a
  discussion of KDE.}  We use {\it Mathematica} functions {\tt
  SmoothHistogram} and {\tt SmoothKernelDistribution} with the default
parameters: smoothing with a Gaussian kernel and an automatic choice
of kernel bandwidth.  Hereafter, in {\it Mathematica} parlance, we
call these ``smooth kernel histograms'' or ``smooth kernel CDFs.''
The smooth kernel distributions capture major differences, if any, by
ignoring spurious detail. 

For semesters 12A--17A, male PIs clearly have better linear ranks than
female PIs.  This is obvious in the histogram plots of the linear rank
distributions, but the CDF and quartile plots better demonstrate their
significance via bootstrapping.  For semesters 17B--21A, the trend is
reversed with female PIs having better linear ranks than male PIs, but
with less significance.  This is only true for linear ranks near the
50$^{\rm th}$ percentile (Q2).  Plots for the remaining 30 data
subsets are shown in the Appendix, Section~\ref{sec:plots}.  For most
semesters and years male PIs have lower linear ranks than female PIs,
but the trend is reversed in some cases.

To help quantify these trends, we list the AD {\it p}-values for the
32 data subsets in Tables~\ref{tab:sem}--\ref{tab:yr}.  The results by
semester are shown in Table~\ref{tab:sem} where we list the semester,
the imbalance key, \adp, both \adpba\ and the 68\% confidence interval
from bootstrap resampling, and the percent of female PIs, SRP members,
and TAC members.  The \adpb\ modeled distributions are not very
Gaussian and therefore \adpba\ is not particularly useful.  The 68\%
confidence interval, however, provides a useful measure of the
uncertainty in \adp.  The horizontal line divides the data into two
parts: before and after we informed our reviewers of the
\citet{lonsdale16} results and started to recruit more female
reviewers. The results for all semesters between 12A--17A and 17B--21A
are given in the last two rows of the table and shown in
Figures~\ref{fig:stats_12a-17a} and \ref{fig:stats_17b-21a}.  Male PIs
are clearly favored over female PIs for the earlier semesters.  After
the administrative changes the trend has reversed to some extent but
with less significance.  The \adp\ numbers are sensitive to small
changes in the data.  This is demonstrated by the large range in some
of the 68\% confidence intervals.  For example, in semester 12A the AD
{\it p}-value is $\rho_{\text{AD}}(x,y) = 0.86$, but the mean AD {\it
  p}-value from bootstrap resampling is
$\rho_{\text{AD}}^{\text{boot}} = 0.34$ with a 68\% confidence
interval of 0.099--0.58.

\begin{deluxetable}{lccccrrr}
\tabletypesize{\scriptsize}
\tablecaption{Gender Results by Semester \label{tab:sem}}
\tablewidth{0pt}
\setlength{\tabcolsep}{2.0pt}
\tablehead{
\colhead{} & \colhead{} & \colhead{} & 
\multicolumn{2}{c}{\underline{~~~~~~AD Bootstrapping~~~}} & 
\multicolumn{3}{c}{\underline{~Percent Female~}} \\ 
\colhead{Semester} & \colhead{IK} & 
\colhead{$\rho_{\text{AD}}(x,y)$} & 
\colhead{$\rho_{\text{AD}}^{\text{boot}}$} & 
\colhead{68\% Conf.} & \colhead{PI} & \colhead{SRP} & \colhead{TAC} 
}
\startdata 
12A & \dots & 0.86 & 0.34 & ($0.099-0.58$) & 26.9 & 25.0 & 12.5 \\ 
12B & \dots & 0.40 & 0.25 & ($0.028-0.50$) & 26.2 & 25.0 & 12.5 \\ 
13A & \dots & 0.34 & 0.21 & ($0.026-0.41$) & 25.5 & 25.0 & 12.5 \\ 
13B & \dots & 0.56 & 0.29 & ($0.042-0.56$) & 24.5 & 27.1 & 25.0 \\ 
14A & \dots & 0.51 & 0.23 & ($0.050-0.43$) & 29.7 & 29.2 & 25.0 \\ 
14B & M & 0.049 & 0.056 & ($0.0030-0.10$) & 30.3 & 25.0 & 25.0 \\ 
15A & \dots & 0.73 & 0.31 & ($0.069-0.56$) & 24.8 & 18.8 & 12.5 \\ 
15B & \dots & 0.54 & 0.28 & ($0.045-0.54$) & 27.0 & 25.0 &  0.0 \\ 
16A & \dots & 0.65 & 0.27 & ($0.065-0.49$) & 30.2 & 25.0 & 12.5 \\ 
16B & M & 0.019 & 0.036 & ($0.00083-0.065$) & 22.6 & 17.0 & 11.1 \\ 
17A & m & 0.12 & 0.15 & ($0.0057-0.32$) & 32.7 & 20.8 & 12.5 \\ 
\tableline 
17B & \dots & 0.25 & 0.13 & ($0.027-0.24$) & 30.6 & 31.5 & 25.0 \\ 
18A & \dots & 0.22 & 0.16 & ($0.017-0.32$) & 32.1 & 38.9 & 25.0 \\ 
18B & F & 0.064 & 0.085 & ($0.0033-0.17$) & 33.6 & 48.1 & 37.5 \\ 
19A & \dots & 0.73 & 0.30 & ($0.061-0.56$) & 32.0 & 48.1 & 50.0 \\ 
19B & \dots & 0.71 & 0.33 & ($0.057-0.62$) & 28.7 & 43.6 & 77.8 \\ 
20A & f & 0.13 & 0.15 & ($0.0048-0.35$) & 36.5 & 39.3 & 66.7 \\ 
20B & \dots & 0.83 & 0.32 & ($0.074-0.57$) & 33.6 & 43.9 & 55.6 \\ 
21A & F & 0.097 & 0.12 & ($0.0037-0.26$) & 31.9 & 43.6 & 55.6 \\ 
 & & & & & & & \\ 
12A--17A & M & 0.0084 & 0.015 & ($0.00038-0.025$) & 27.3 & 23.9 & 14.6 \\ 
17B$-$21A & f & 0.11 & 0.11 & ($0.0057-0.24$) & 32.5 & 42.1 & 49.1 \\ 
\enddata 
  \tablecomments{There are typically about 300 proposals per semester.} 
\end{deluxetable}

\begin{deluxetable}{lcccccccrrr}
\tabletypesize{\scriptsize}
\tablecaption{Gender Results by Year \label{tab:yr}}
\tablewidth{0pt}
\setlength{\tabcolsep}{2.0pt}
\tablehead{
\colhead{} & \multicolumn{4}{c}{\underline{~~~~~Imbalance Key~~~~~}} & \colhead{} & 
\multicolumn{2}{c}{\underline{~~~~~~AD Bootstrapping~~~}} & 
\multicolumn{3}{c}{\underline{~Percent Female~}} \\ 
\colhead{Year} & \colhead{GBT} & \colhead{VLA} & \colhead{VLBA} & \colhead{All} & 
\colhead{$\rho_{\text{AD}}(x,y)$} & 
\colhead{$\rho_{\text{AD}}^{\text{boot}}$} & 
\colhead{68\% Conf.} & \colhead{PI} & \colhead{SRP} & \colhead{TAC} 
}
\startdata 
2012 & \dots & \dots & \dots & \dots & 0.39 & 0.21 & ($0.033-0.41$) & 26.6 & 25.0 & 12.5 \\ 
2013 & F & \dots & \dots & \dots & 0.20 & 0.18 & ($0.011-0.38$) & 25.0 & 26.0 & 18.8 \\ 
2014 & M & \dots & \dots & M & 0.040 & 0.046 & ($0.0025-0.085$) & 30.0 & 27.1 & 25.0 \\ 
2015 & \dots & \dots & \dots & \dots & 0.71 & 0.28 & ($0.073-0.49$) & 25.7 & 21.9 &  6.2 \\ 
2016 & \dots & m & \dots & m & 0.13 & 0.11 & ($0.0086-0.22$) & 26.5 & 21.0 & 11.8 \\ 
2017 & \dots & \dots & \dots & M & 0.086 & 0.087 & ($0.0047-0.17$) & 31.6 & 26.2 & 18.8 \\ 
2018 & m & f & \dots & \dots & 0.23 & 0.13 & ($0.023-0.23$) & 32.7 & 43.5 & 31.2 \\ 
2019 & \dots & \dots & \dots & \dots & 0.68 & 0.32 & ($0.048-0.63$) & 30.8 & 45.9 & 63.9 \\ 
2020 & \dots & \dots & \dots & \dots & 0.61 & 0.28 & ($0.044-0.54$) & 35.0 & 41.6 & 61.1 \\ 
2021 & f & \dots & \dots & F & 0.097 & 0.12 & ($0.0041-0.25$) & 31.9 & 43.6 & 55.6 \\ 
 & & & & & & & & & & \\ 
All & \dots & f & \dots & M & 0.031 & 0.029 & ($0.0022-0.052$) & 29.3 & 43.6 & 29.2 \\ 
\enddata 
  \tablecomments{Statistics for year 2021 only includes semester A.} 
\end{deluxetable}

Following \citet{lonsdale16}, we show the results by year by averaging
over two semesters (see Table~\ref{tab:yr}).  This increases the
sensitivity and allows us to show the imbalance key by telescope.  The
smaller number of proposals submitted to the VLBA, however, is still
not statistically significant and therefore this column is blank and
only included for completeness.  Inspection of the imbalance keys
shows the transition from male to female after the administrative
changes.

\section{Discussion}\label{sec:discussion}

There is mounting evidence of systematic, gender-related trends in the
review of astronomical observing proposals.  This was first discovered
in HST observing proposals where male PIs had better success rates
than female PIs \citep{reid14}.  Similar trends have been found at ESO
\citep{patat16}, ALMA \citep{lonsdale16, carpenter20}, NRAO
\citep{lonsdale16}, and the Canada-France-Hawaii Telescope (CFHT) and
Gemini Observatory \citep{spekkens18}.

Unconscious bias may be the cause of the measured gender imbalance.
But unconscious bias is complex, unintended, and multifaceted.  Other
effects such as prestige bias complicate the interpretation of the
data.  For example, some studies have noted a dependence on the
gender-based success rate and PI seniority \citep{spekkens18}; the
seniority of the review panels \citep{patat16}; and the potential
privilege of a proposer being a member of the TAC
\citep{greaves18}---although \citet{carpenter20} did not find any
correlation.  Numerous studies have noted that both males and females
evaluate work by males higher than females for similar results. This
bias has been discussed in several professional fields \citep[see][and
references within]{spekkens18, johnson20}.  This would lead to the
conclusion that changing the gender distribution of the review panels
would not by itself rectify the situation.  \citet{lonsdale16} found
no evidence that the panel gender composition was causing the gender
imbalance, but here we do see a correlation between the gender
distribution on the review panels and linear ranks.  But we also
started briefing the SRPs about the gender imbalance at the same time
and therefore we cannot associate this correlation with the improved
female PI linear ranks.

The STScI has adopted a dual-anonymous proposal review process for the
HST where the identity of both the authors and reviewers are hidden.
This requires that the proposal be anonymized and thus proposers have
to consciously remove any information in the proposal that points to
their team.  Many other observatories are following this approach
(e.g., ALMA, ESO, most NASA programs, etc.).  \citet{johnson20} argue
that unconscious bias is difficult to overcome.  Many of the
procedures used to combat unconscious bias, such as training programs,
are not effective.  Moreover, there can be a backlash of a perceived
advantage.  So instead of ``fixing'' the problem, a dual-anonymous
approach reduces the likelihood of a bias from entering into the
process.  For these reasons the NRAO has decided to adopt a
dual-anonymous proposal review process for the new telescope time
allocation software currently being developed.

\section{Conclusions}\label{sec:conclusions}

Studies reveal gender-related systematic trends in the peer review of
observing proposals from astronomical observatories \citep{reid14,
  patat16, lonsdale16, spekkens18}.  The exact cause of the measured
gender imbalance is not certain and in some cases there are other
factors at play (e.g., seniority).  Here we investigate gender
systematic effects from semesters 12A--21A in AUI's NA facilities:
VLA, VLBA, and GBT.  We use the linear ranks that are finalized after
the SRP has formed their consensus review.  To interpret these data we
produce histograms, CDFs, and the quartile values of the male and
female PI linear ranks.  We use bootstrap resampling to assess the
uncertainty in the CDFs and to produce a distribution of quartile
values.  To determine if the distribution of male and female PI linear
ranks are from the same parent distribution, we calculate the
Anderson-Darling statistic.

Over all semesters we find that male PIs are favored over female PIs
with $\rho_{\text{AD}}(x,y) = 0.031$.  The 68\% confidence interval
via bootstrap resampling is 0.0022--0.052.  Starting in semester 17B,
the NRAO made two administrative changes: (1) we informed SRP and TAC
members of the gender imbalance found by \citet{lonsdale16}; and (2)
we increased the female membership on the SRP and TAC to reflect the
community demographics.  These changes appear to have ameliorated the
gender imbalance. Between semesters 12A--17A male PIs are favored over
female PIs with $\rho_{\text{AD}}(x,y) = 0.0084$, whereas between
semesters 17B--21A female PIs are slightly favored over male PIs with
$\rho_{\text{AD}}(x,y) = 0.11$.  The gender imbalance may have merely
been reversed, however, but at a lower significance.

Many observatories have adopted a dual-anonymous approach to proposal
review to reduce the possibility of a bias, instead of trying to
control biases with training or other methods.  The NRAO plans to
adopt dual-anonymous review with the new software being developed for
our telescope time allocation system.

\acknowledgements

We thank the following people for constructive comments on the
manuscript: John Carpenter, Allison Costa, Jeff Kern, Adele Plunkett,
and Lyndele von Schill.  The National Radio Astronomy Observatory is a
facility of the National Science Foundation operated under cooperative
agreement by Associated Universities, Inc.

\software{{\it Mathematica} \citep{wolfram21}.}

\clearpage

\appendix

\section{Statistical Plots}\label{sec:plots}

\begin{figure}
  \centering
  \includegraphics[angle=0,scale=0.9]{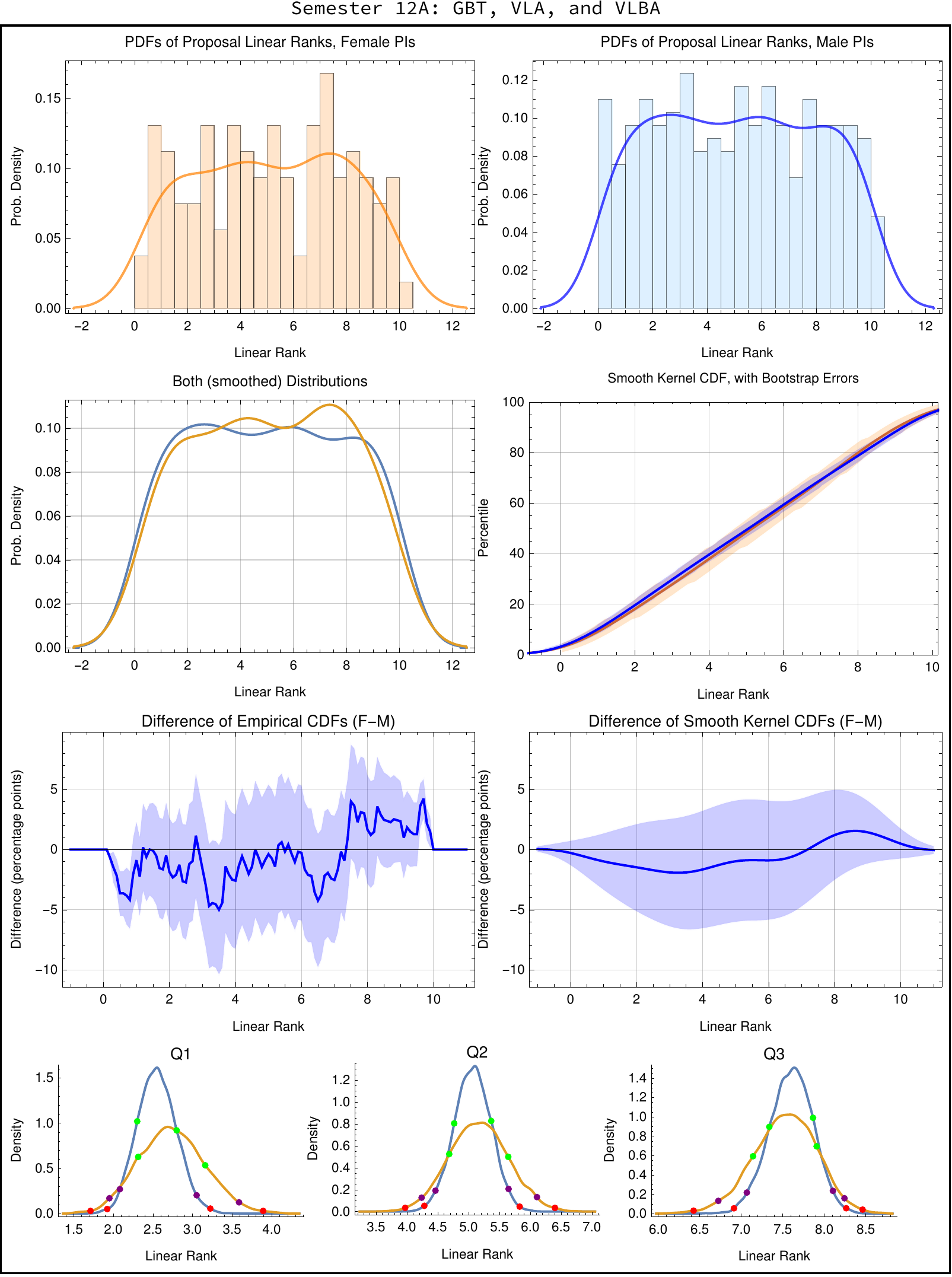}
  \caption{Statistics for semester 12A\@. See
    Figure~\ref{fig:stats_12a-17a} for details.}
\label{fig:stats_12a}
\end{figure}

\begin{figure}
  \centering
  \includegraphics[angle=0,scale=0.9]{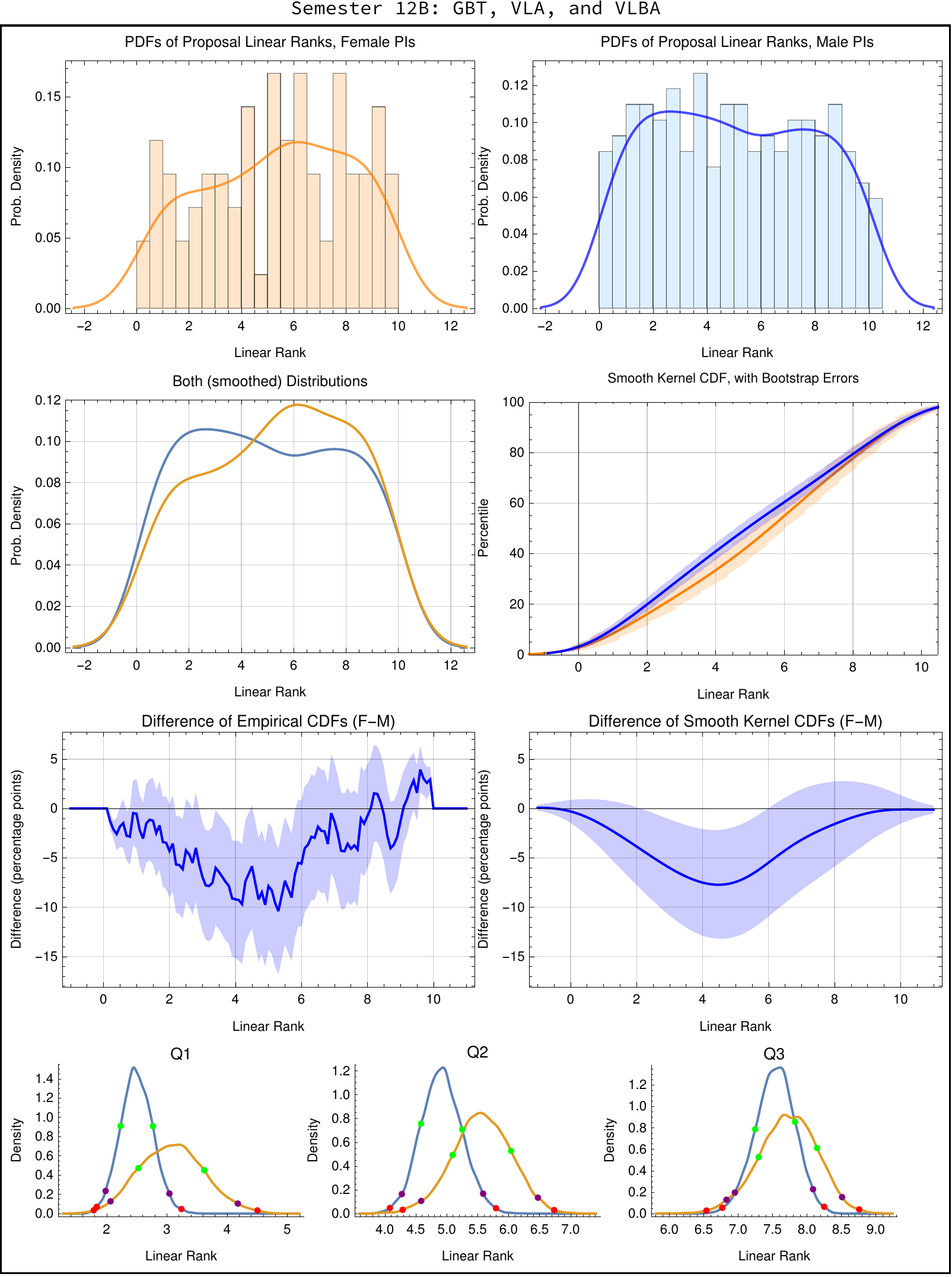}
  \caption{Statistics for semesters 12B\@. See
    Figure~\ref{fig:stats_12a-17a} for details.}
\label{fig:stats_12b}
\end{figure}

\begin{figure}
  \centering
  \includegraphics[angle=0,scale=0.9]{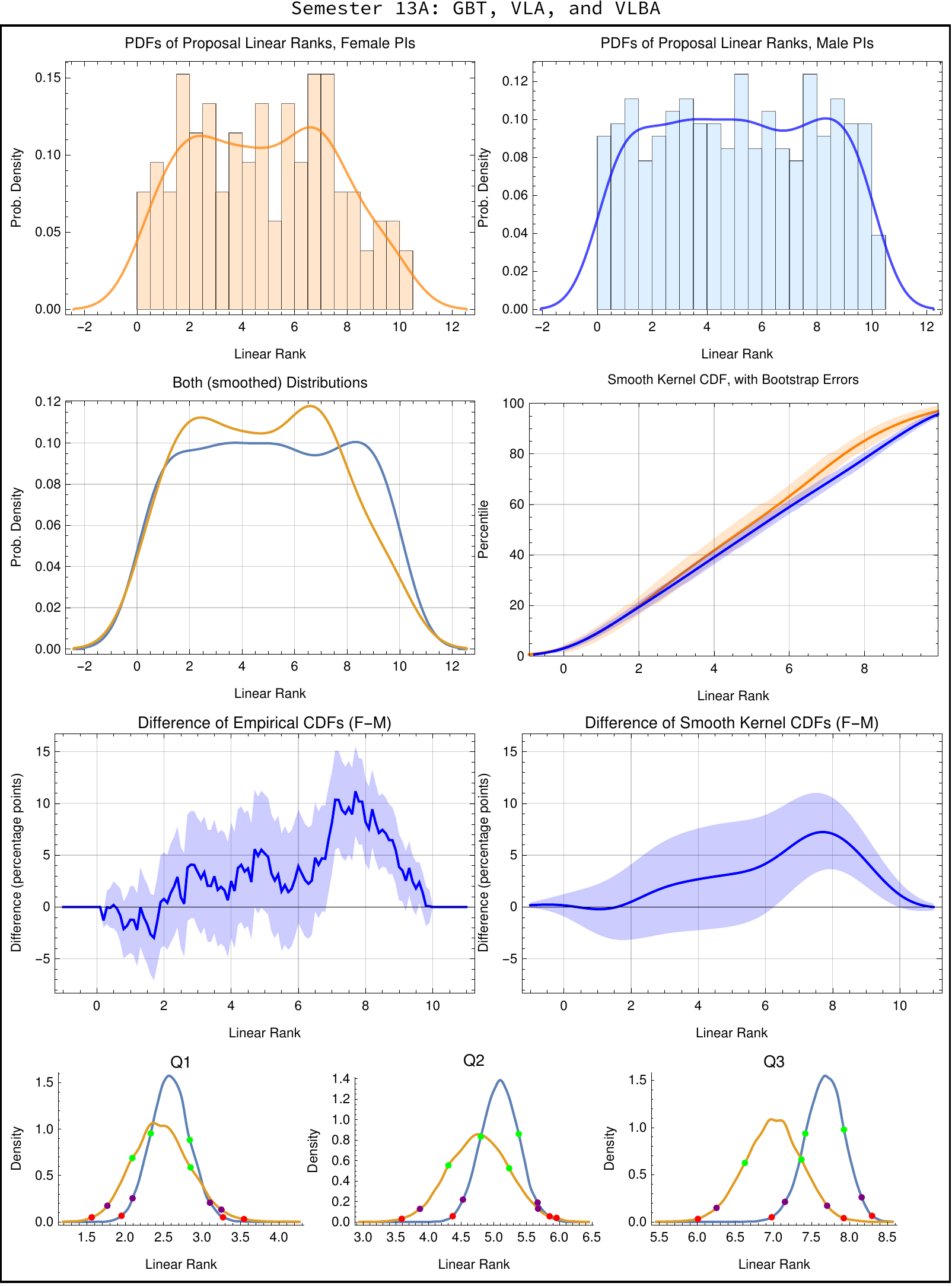}
  \caption{Statistics for semester 13A\@. See
    Figure~\ref{fig:stats_12a-17a} for details.}
\label{fig:stats_13a}
\end{figure}

\begin{figure}
  \centering
  \includegraphics[angle=0,scale=0.9]{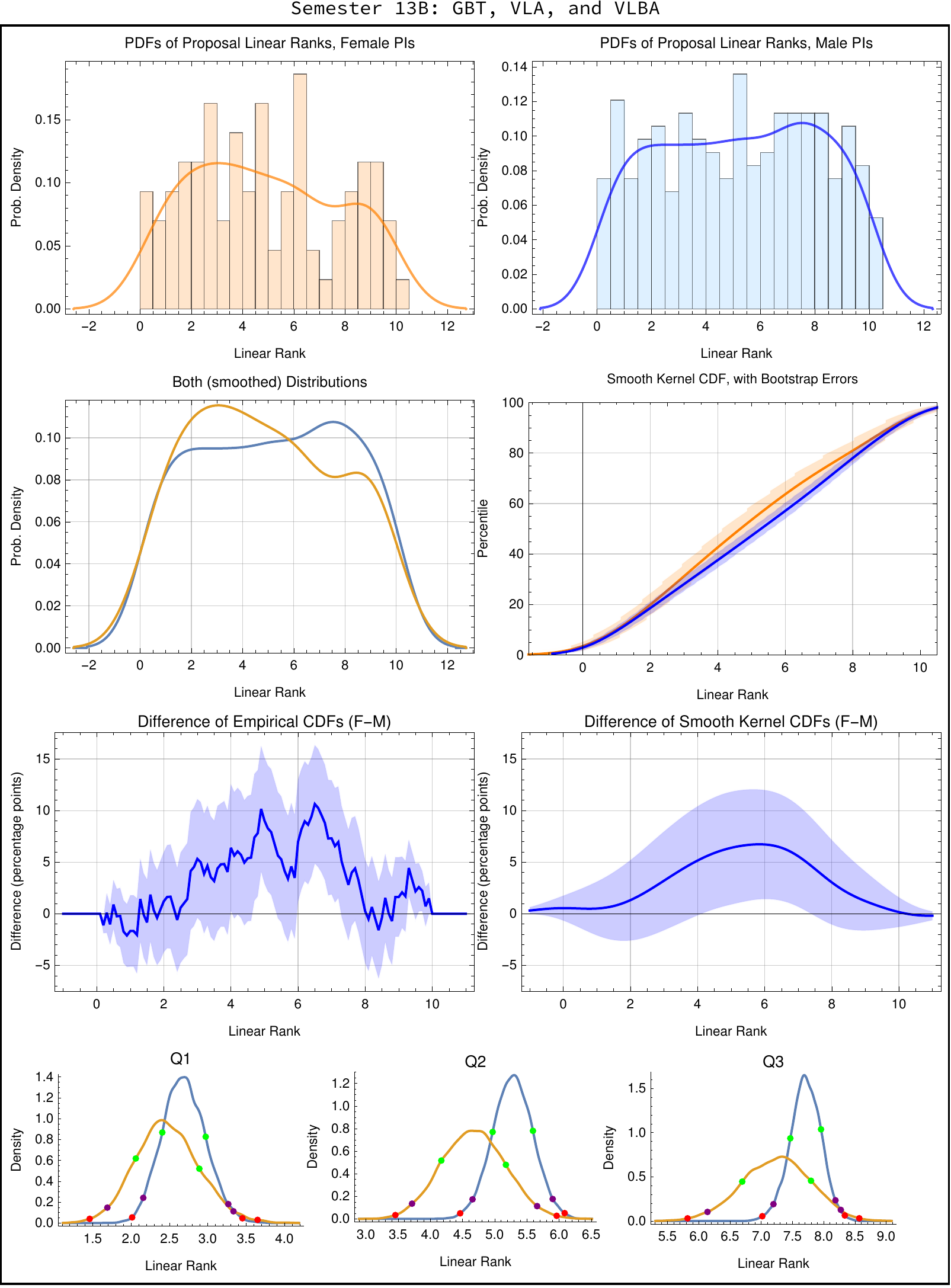}
  \caption{Statistics for semester 13B\@. See
    Figure~\ref{fig:stats_12a-17a} for details.}
\label{fig:stats_13b}
\end{figure}

\begin{figure}
  \centering
  \includegraphics[angle=0,scale=0.9]{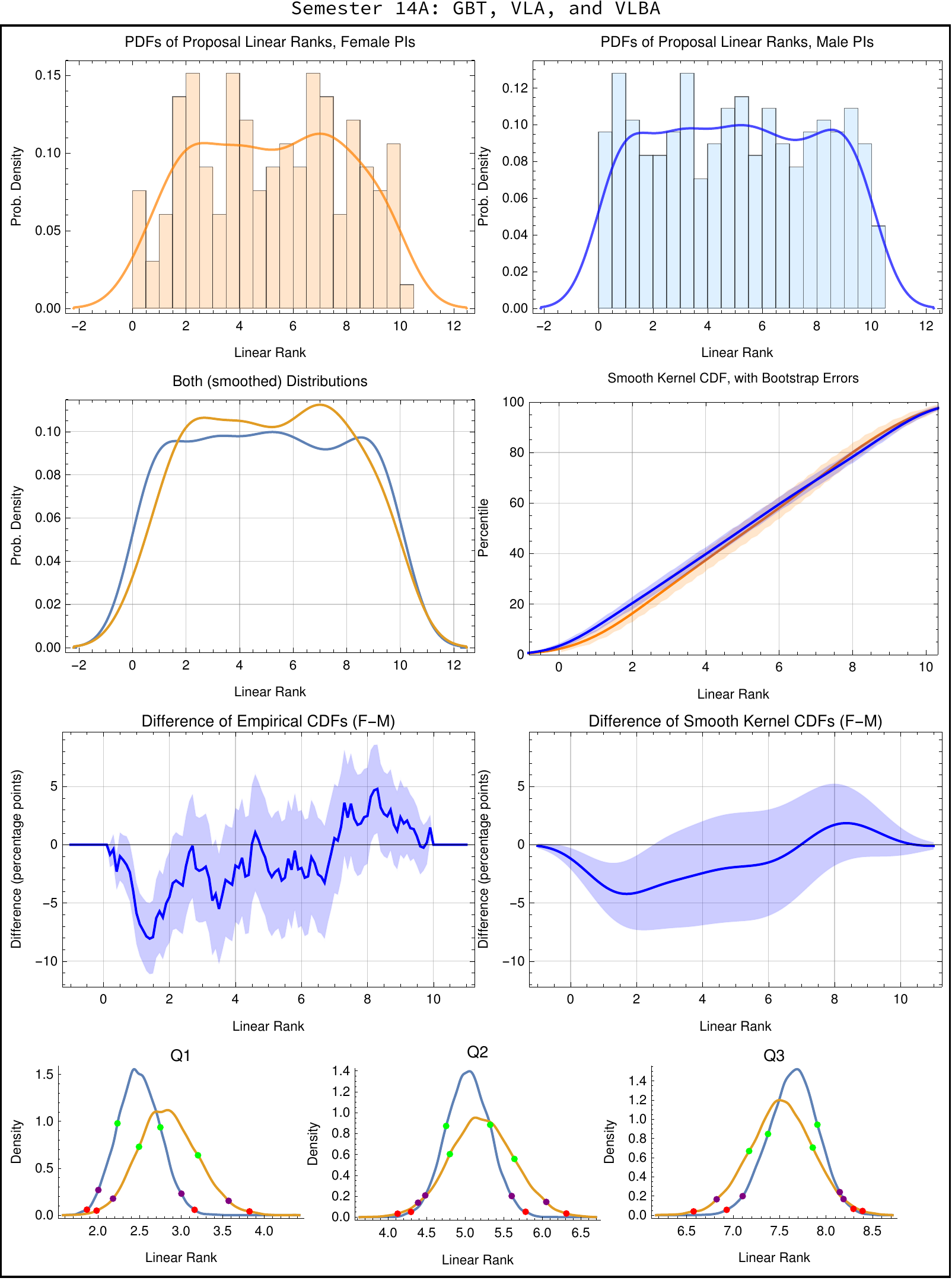}
  \caption{Statistics for semester 14A\@. See
    Figure~\ref{fig:stats_12a-17a} for details.}
\label{fig:stats_14a}
\end{figure}

\begin{figure}
  \centering
  \includegraphics[angle=0,scale=0.9]{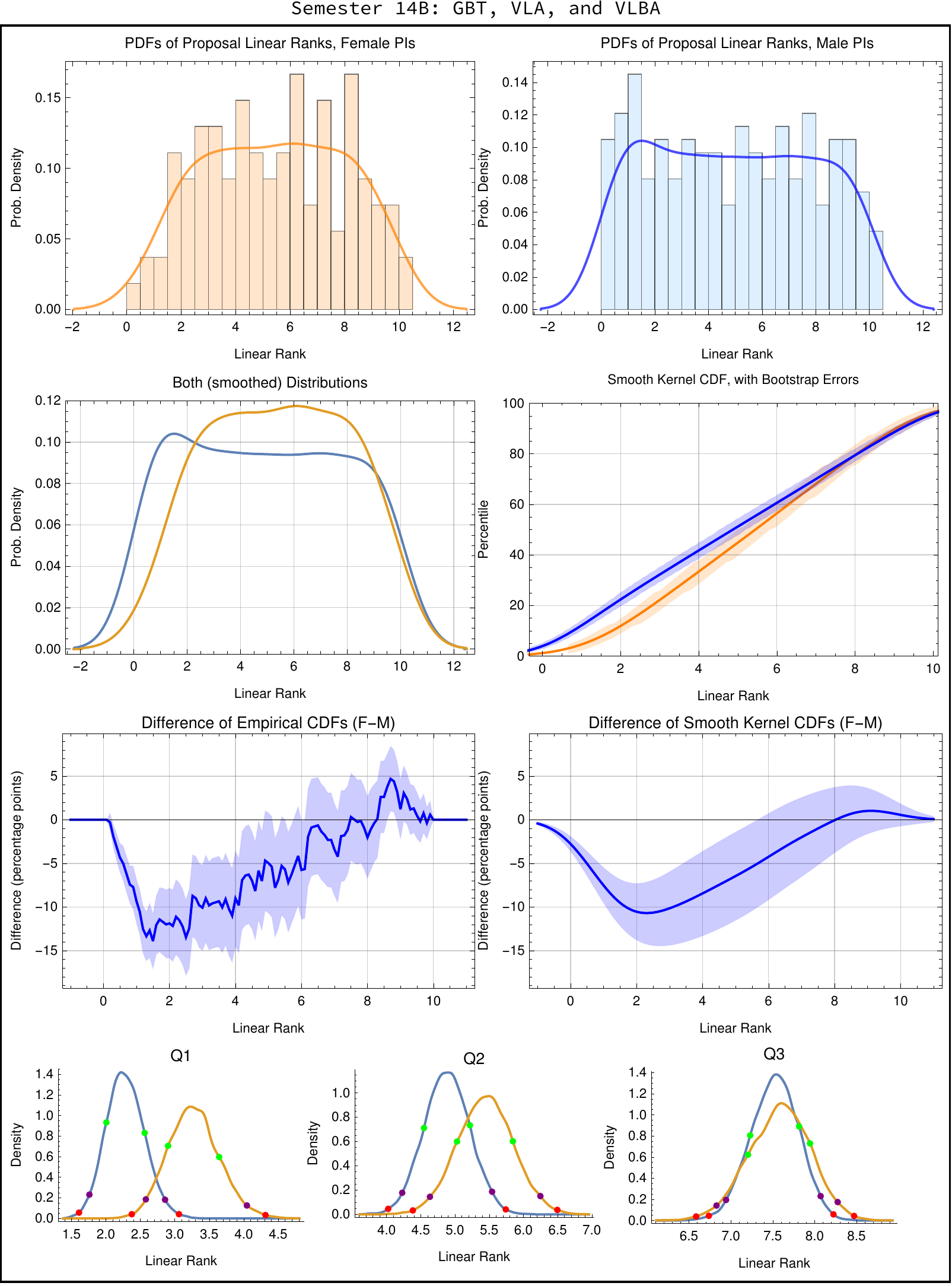}
  \caption{Statistics for semester 14B\@. See
    Figure~\ref{fig:stats_12a-17a} for details.}
\label{fig:stats_14b}
\end{figure}

\begin{figure}
  \centering
  \includegraphics[angle=0,scale=0.9]{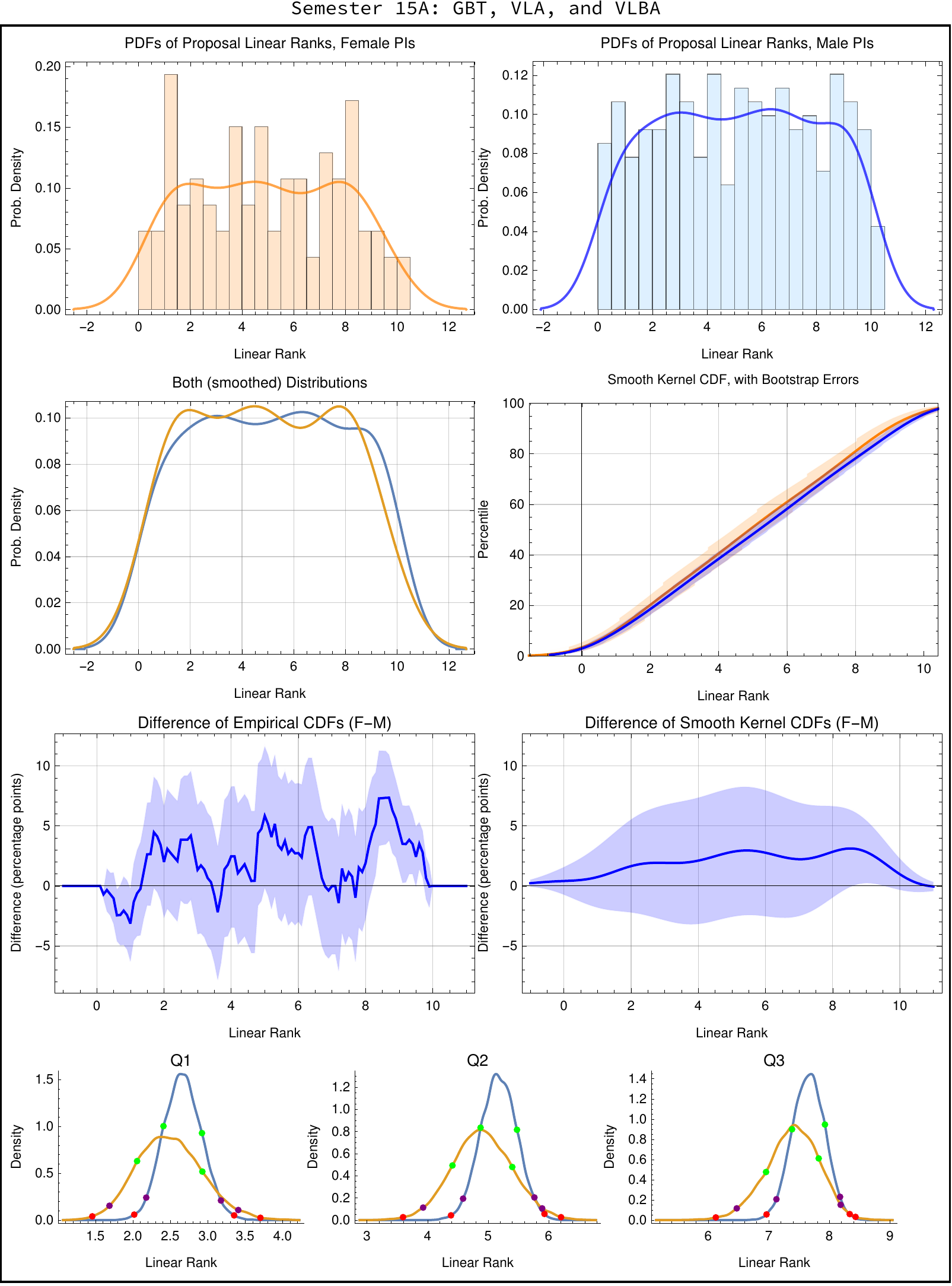}
  \caption{Statistics for semester 15A\@. See
    Figure~\ref{fig:stats_12a-17a} for details.}
\label{fig:stats_15a}
\end{figure}

\begin{figure}
  \centering
  \includegraphics[angle=0,scale=0.9]{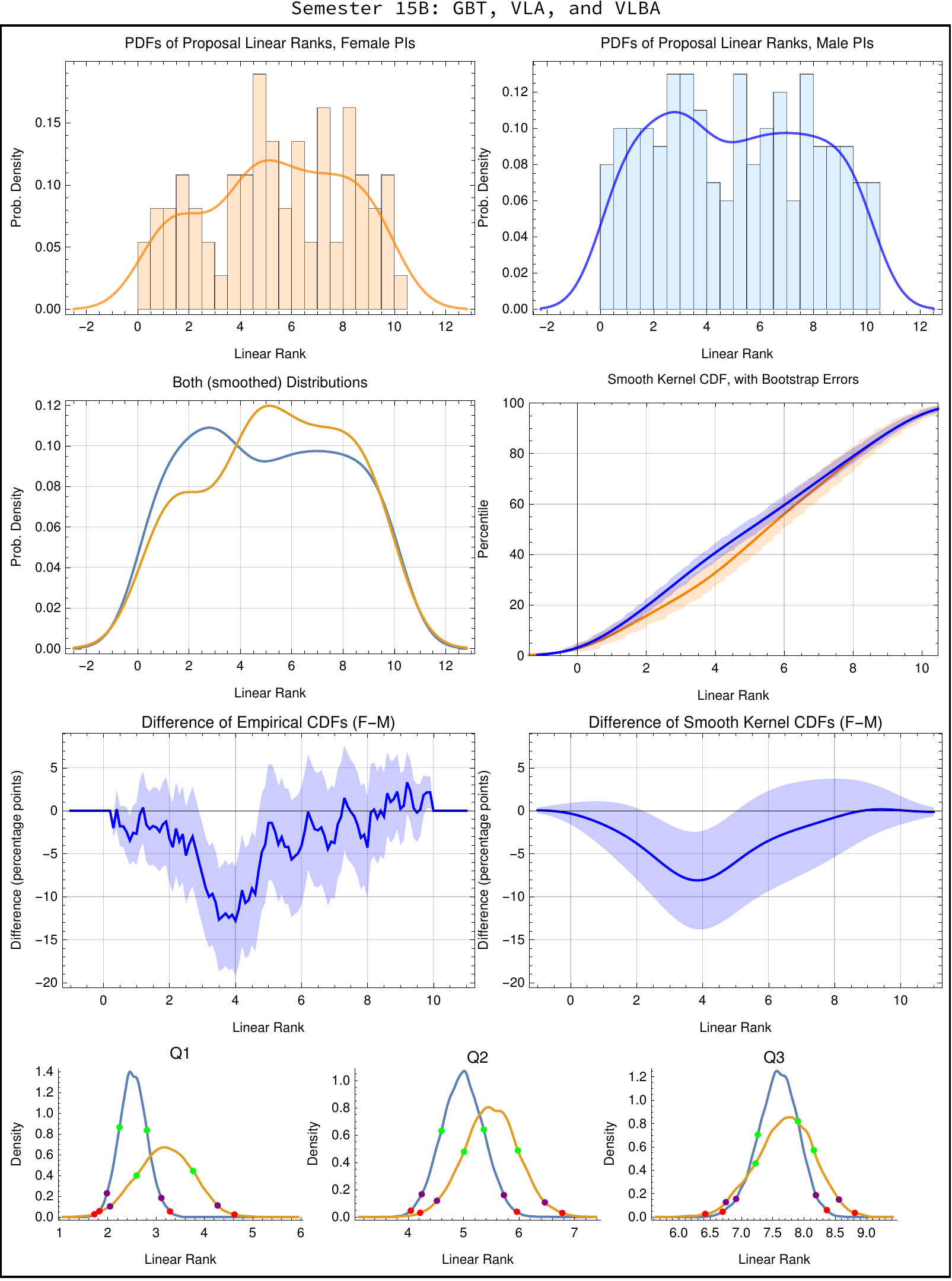}
  \caption{Statistics for semester 15B\@. See
    Figure~\ref{fig:stats_12a-17a} for details.}
\label{fig:stats_15b}
\end{figure}

\begin{figure}
  \centering
  \includegraphics[angle=0,scale=0.9]{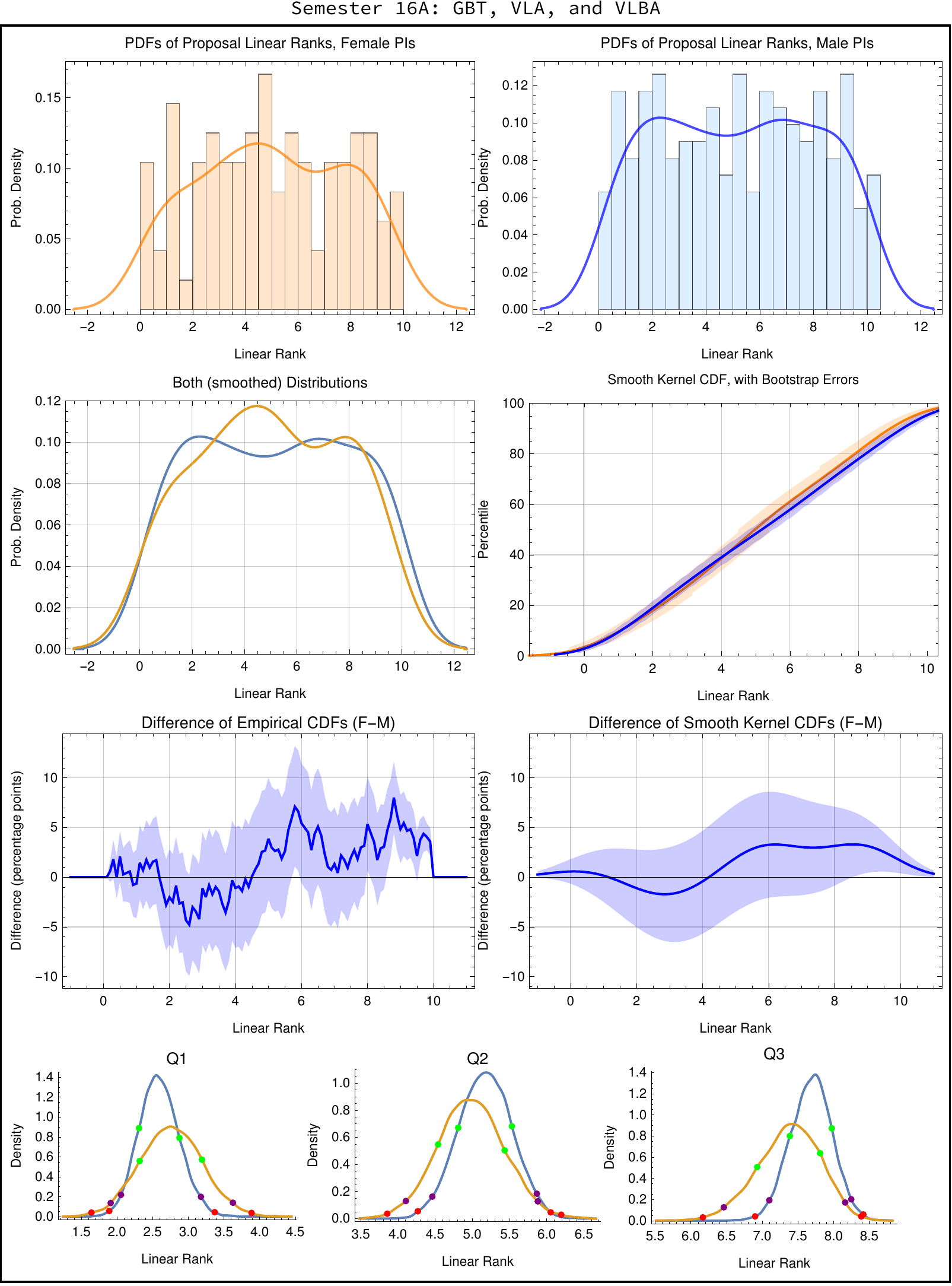}
  \caption{Statistics for semester 16A\@. See
    Figure~\ref{fig:stats_12a-17a} for details.}
\label{fig:stats_16a}
\end{figure}

\begin{figure}
  \centering
  \includegraphics[angle=0,scale=0.9]{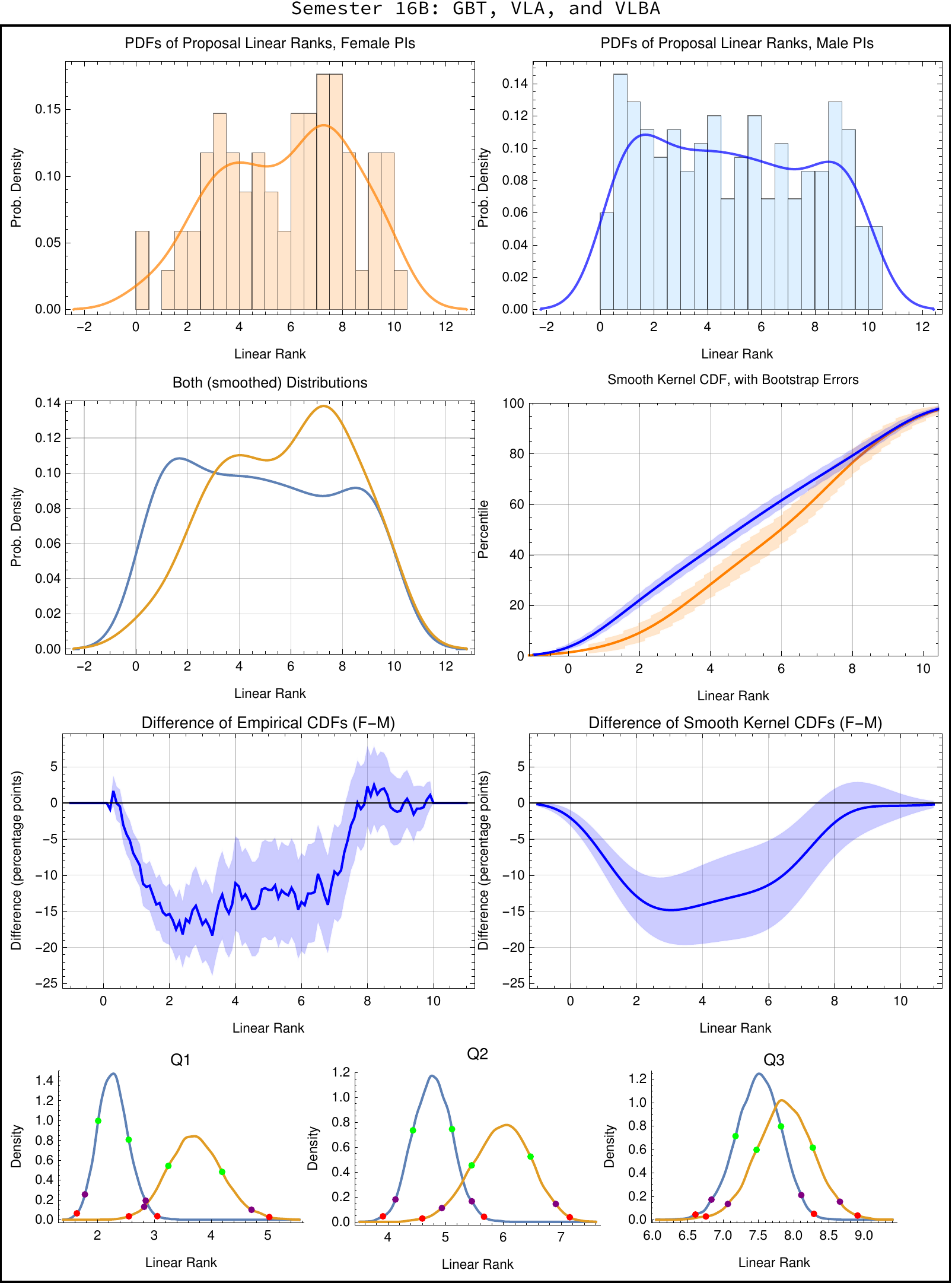}
  \caption{Statistics for semester 16B\@. See
    Figure~\ref{fig:stats_12a-17a} for details.}
\label{fig:stats_16b}
\end{figure}

\begin{figure}
  \centering
  \includegraphics[angle=0,scale=0.9]{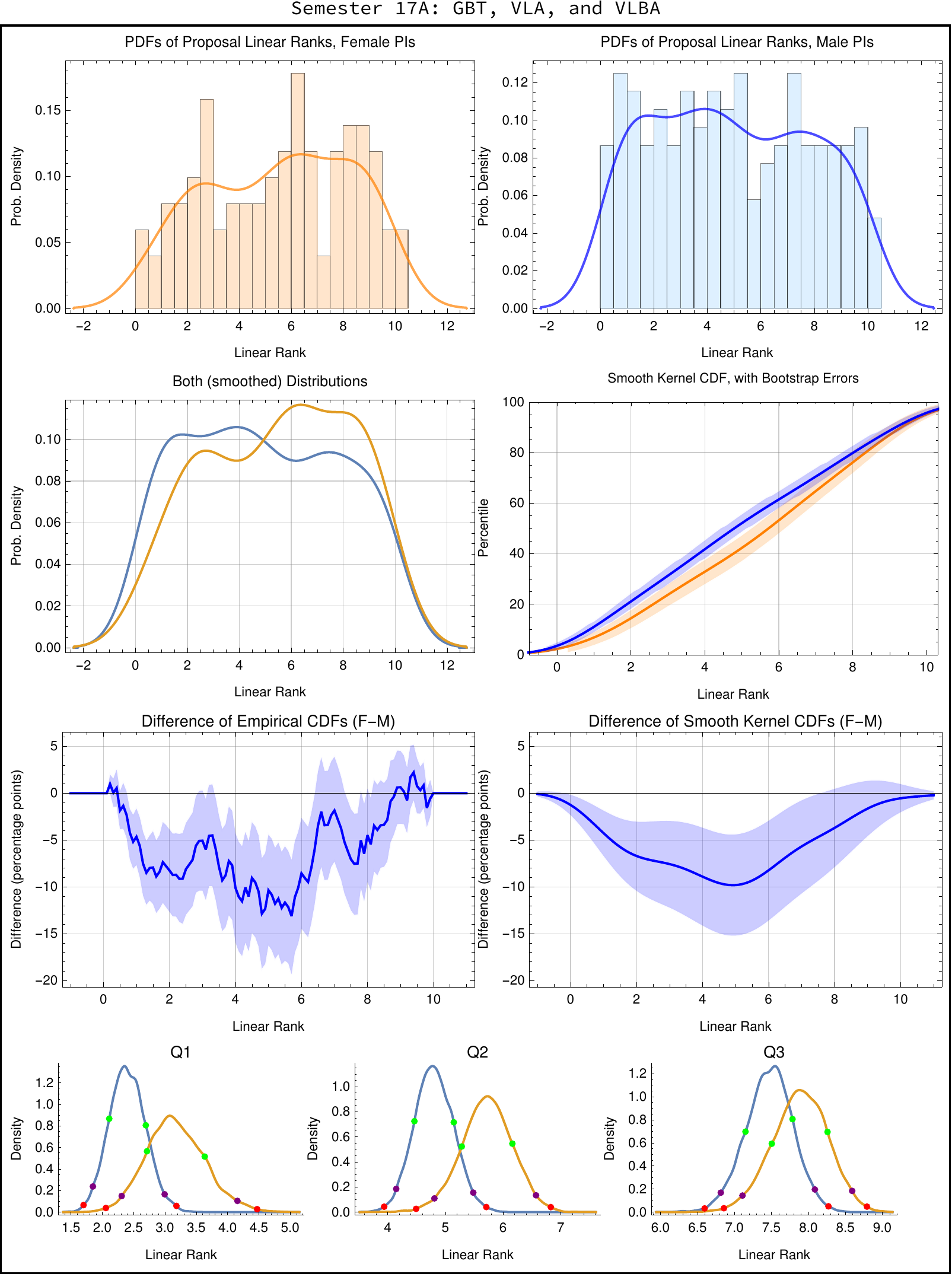}
  \caption{Statistics for semester 17A\@. See
    Figure~\ref{fig:stats_12a-17a} for details.}
\label{fig:stats_17a}
\end{figure}

\begin{figure}
  \centering
  \includegraphics[angle=0,scale=0.9]{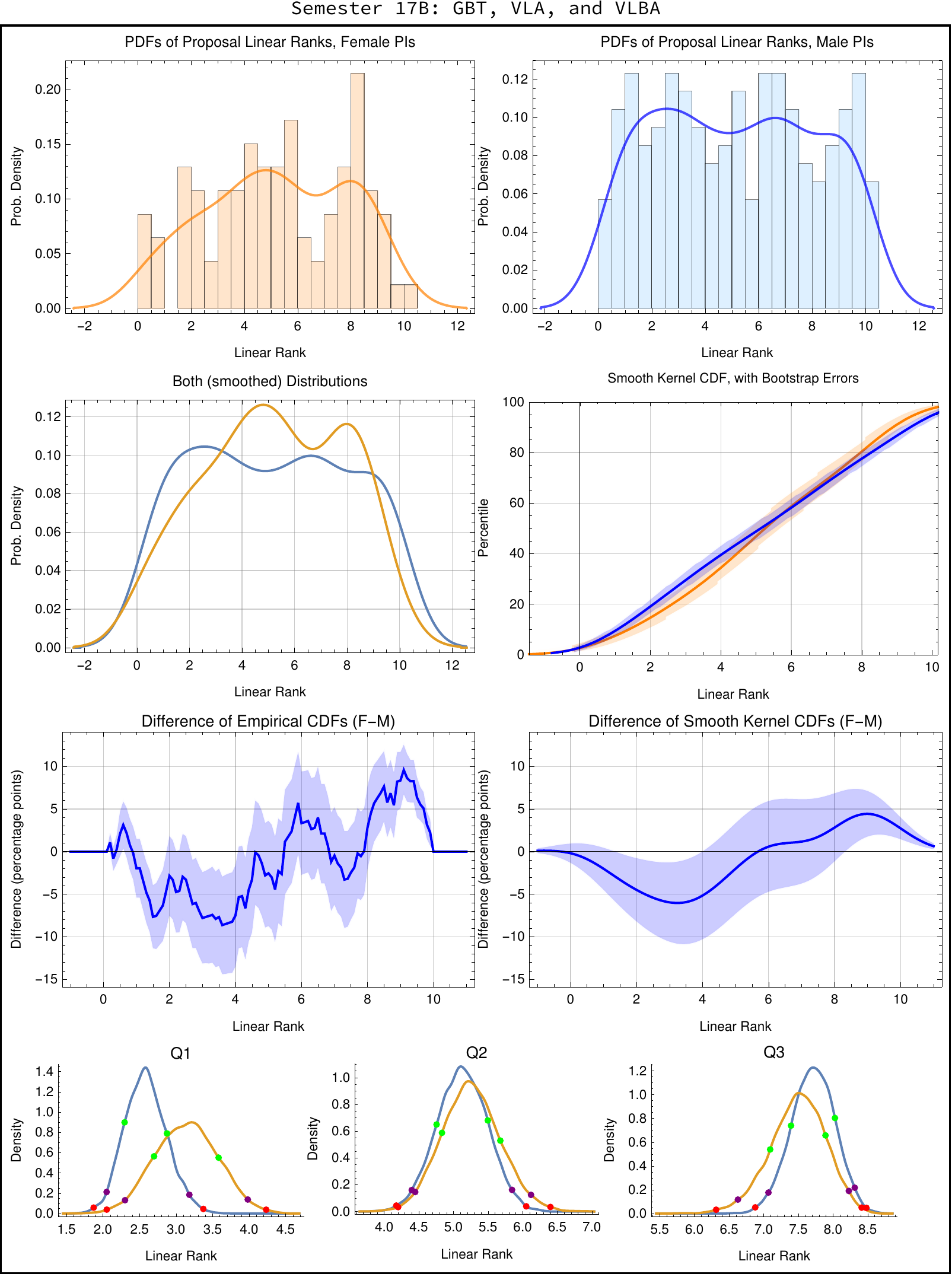}
  \caption{Statistics for semester 17B\@. See
    Figure~\ref{fig:stats_12a-17a} for details.}
\label{fig:stats_17b}
\end{figure}

\begin{figure}
  \centering
  \includegraphics[angle=0,scale=0.9]{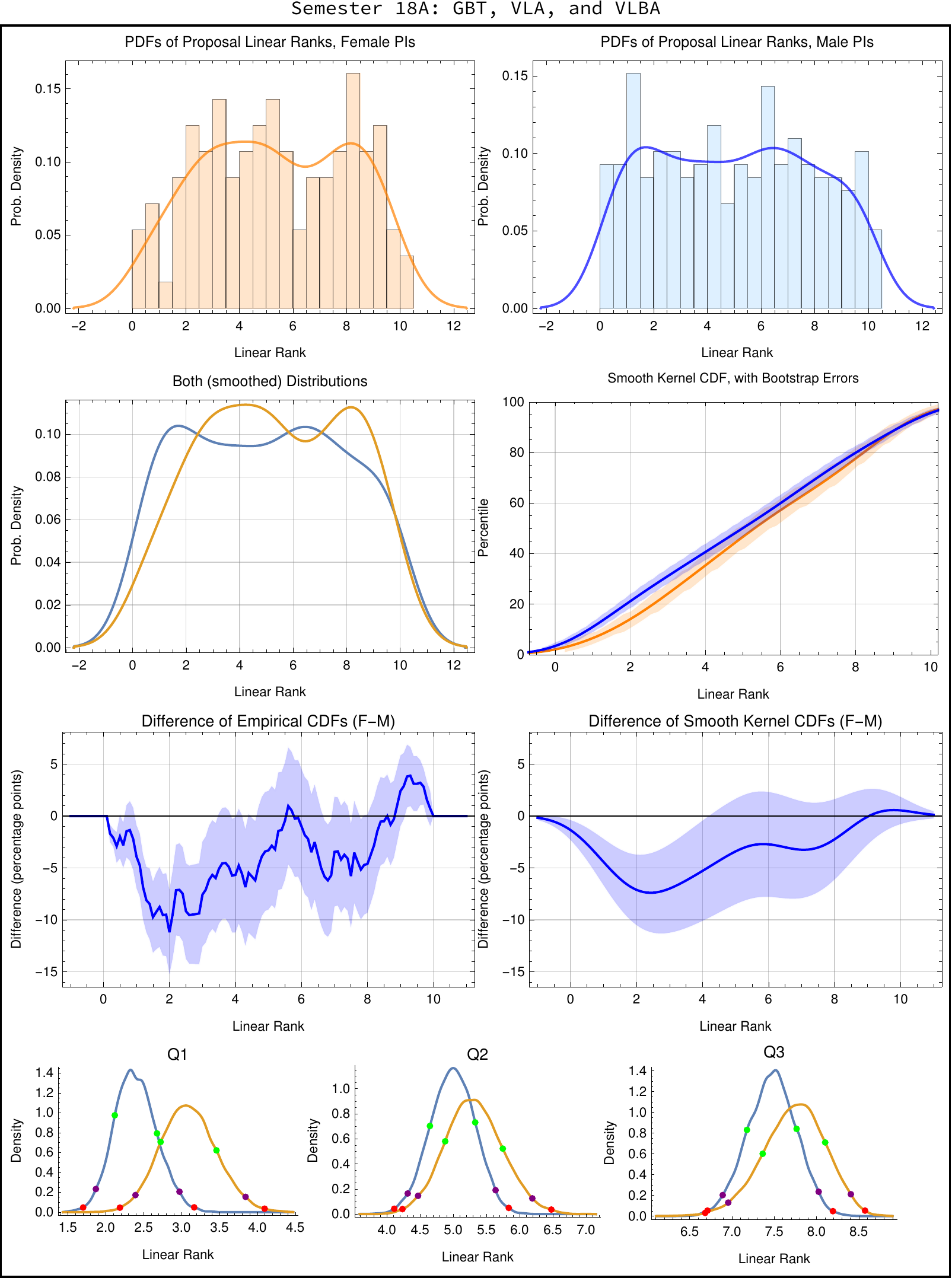}
  \caption{Statistics for semester 18A\@. See
    Figure~\ref{fig:stats_12a-17a} for details.}
\label{fig:stats_18a}
\end{figure}

\begin{figure}
  \centering
  \includegraphics[angle=0,scale=0.9]{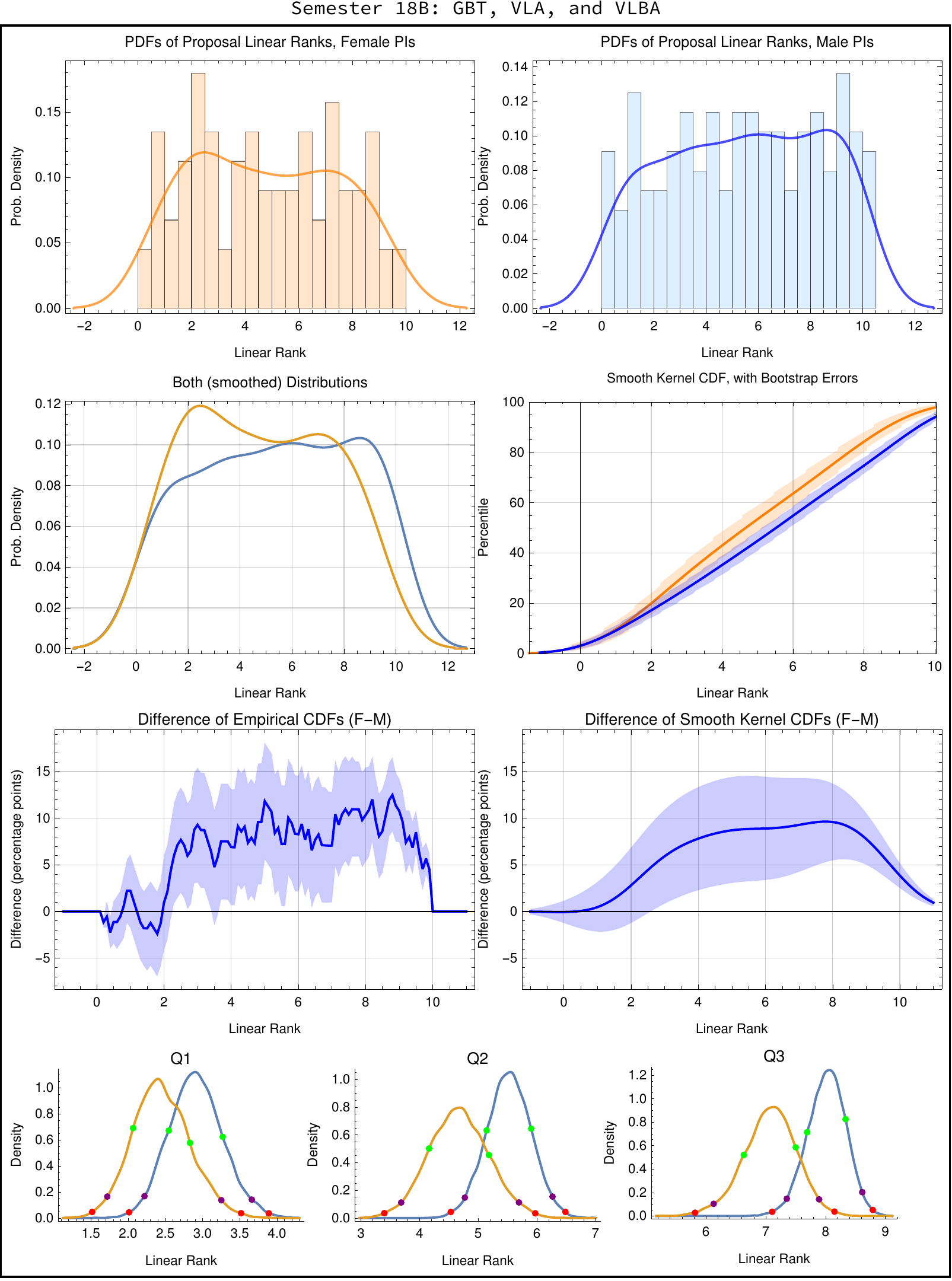}
  \caption{Statistics for semester 18B\@. See
    Figure~\ref{fig:stats_12a-17a} for details.}
\label{fig:stats_18b}
\end{figure}

\begin{figure}
  \centering
  \includegraphics[angle=0,scale=0.9]{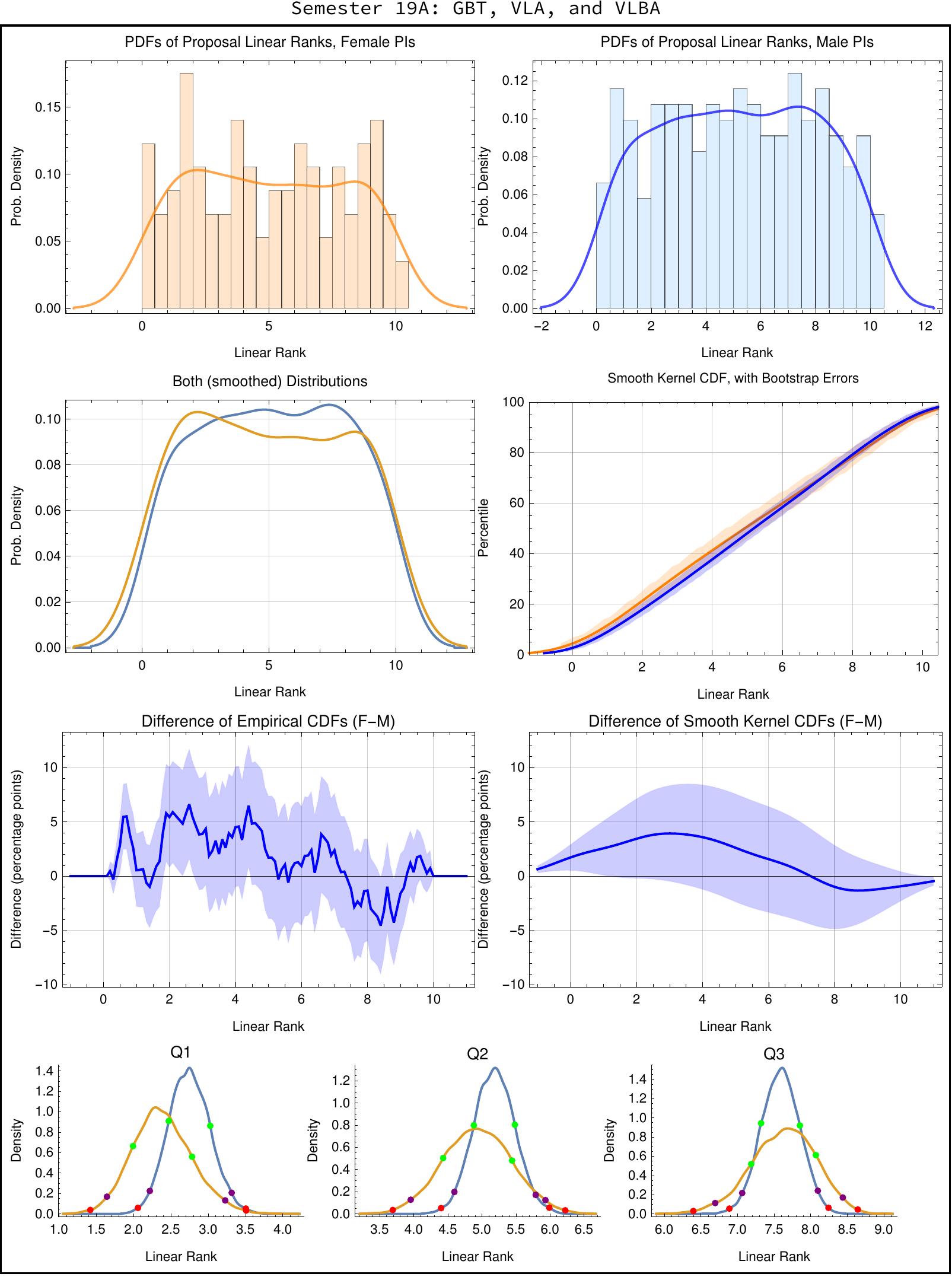}
  \caption{Statistics for semester 19A\@. See
    Figure~\ref{fig:stats_12a-17a} for details.}
\label{fig:stats_19a}
\end{figure}

\begin{figure}
  \centering
  \includegraphics[angle=0,scale=0.9]{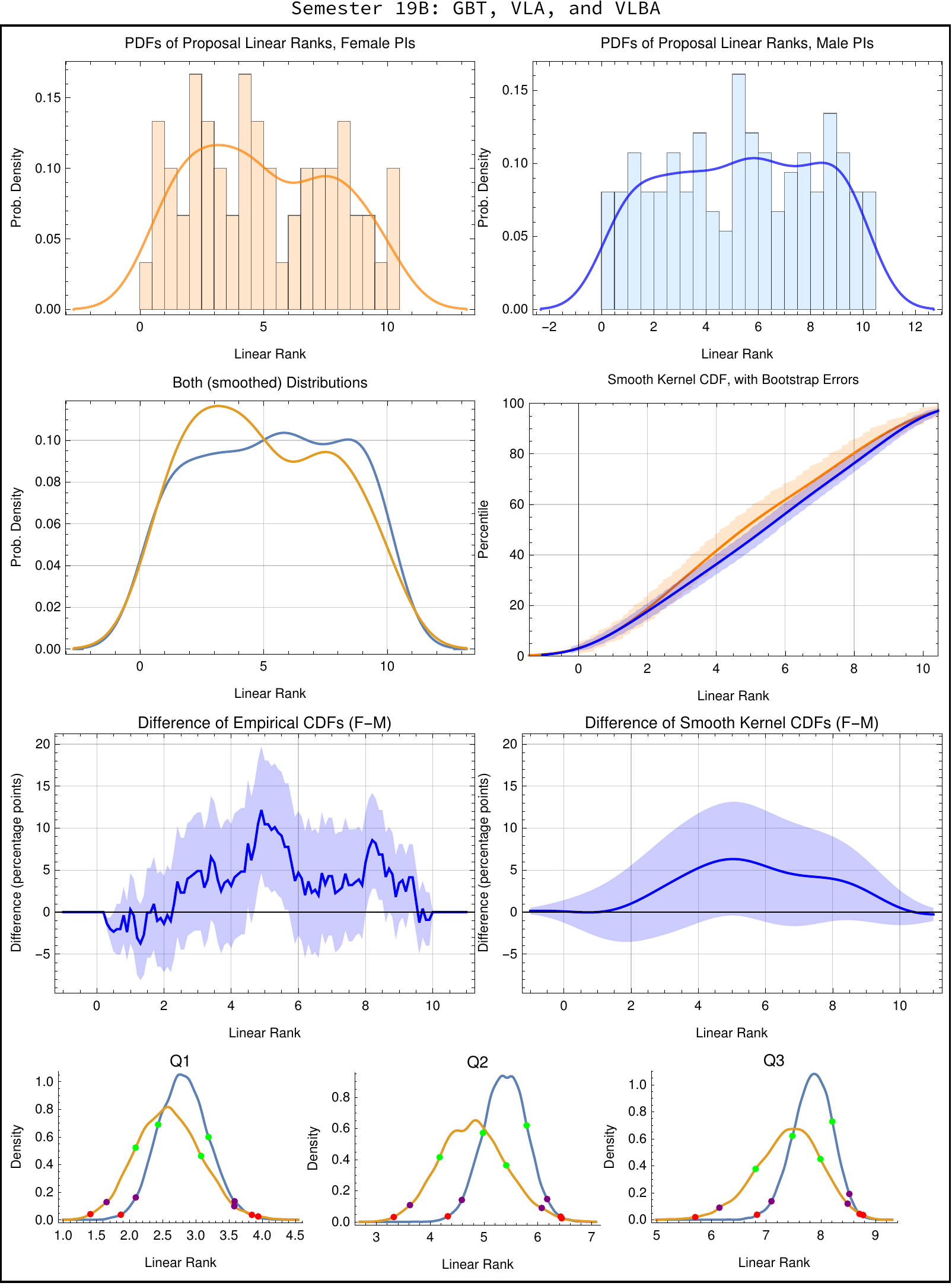}
  \caption{Statistics for semester 19B\@. See
    Figure~\ref{fig:stats_12a-17a} for details.}
\label{fig:stats_19b}
\end{figure}

\begin{figure}
  \centering
  \includegraphics[angle=0,scale=0.9]{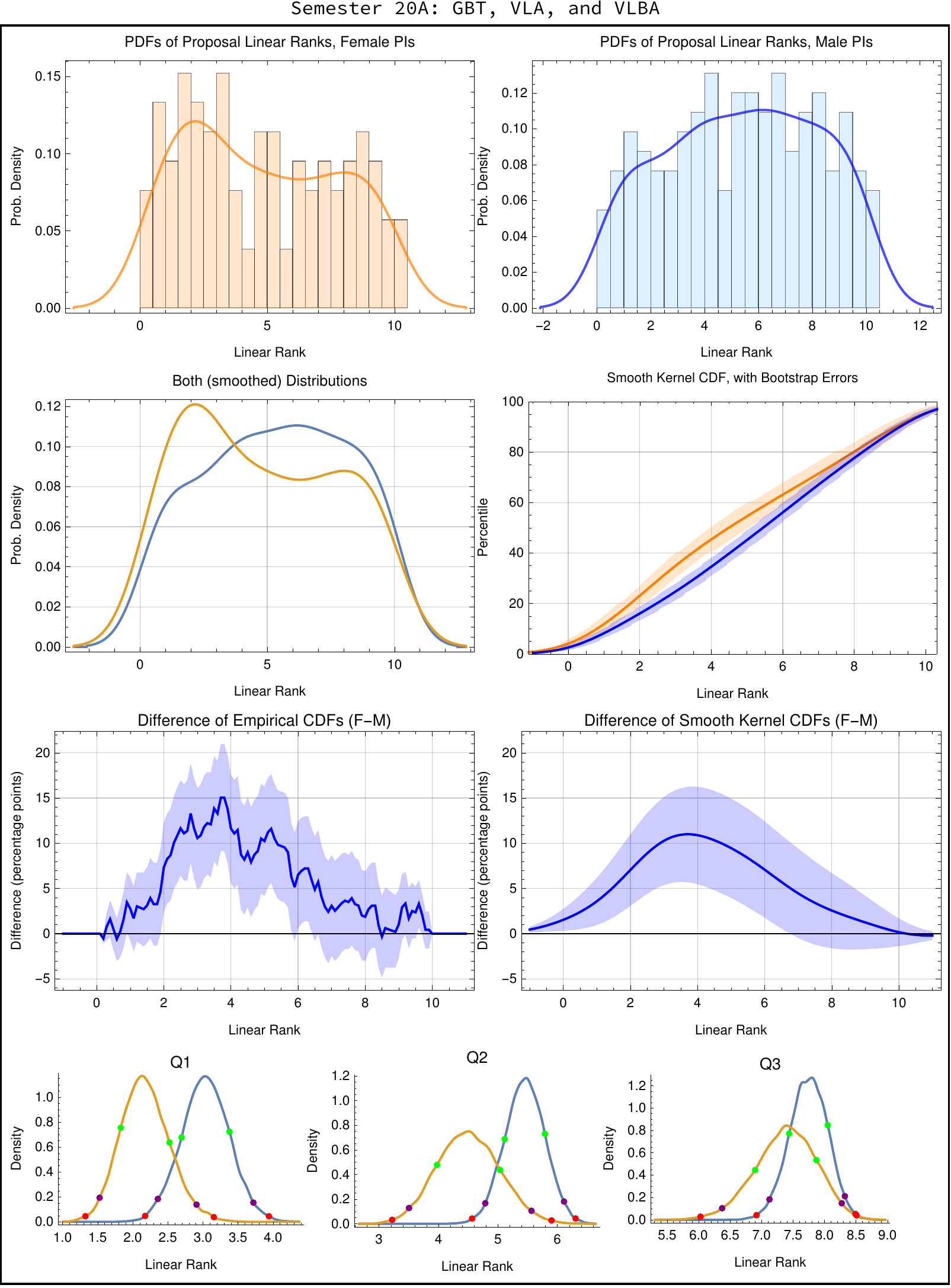}
  \caption{Statistics for semester 20A\@. See
    Figure~\ref{fig:stats_12a-17a} for details.}
\label{fig:stats_20a}
\end{figure}

\begin{figure}
  \centering
  \includegraphics[angle=0,scale=0.9]{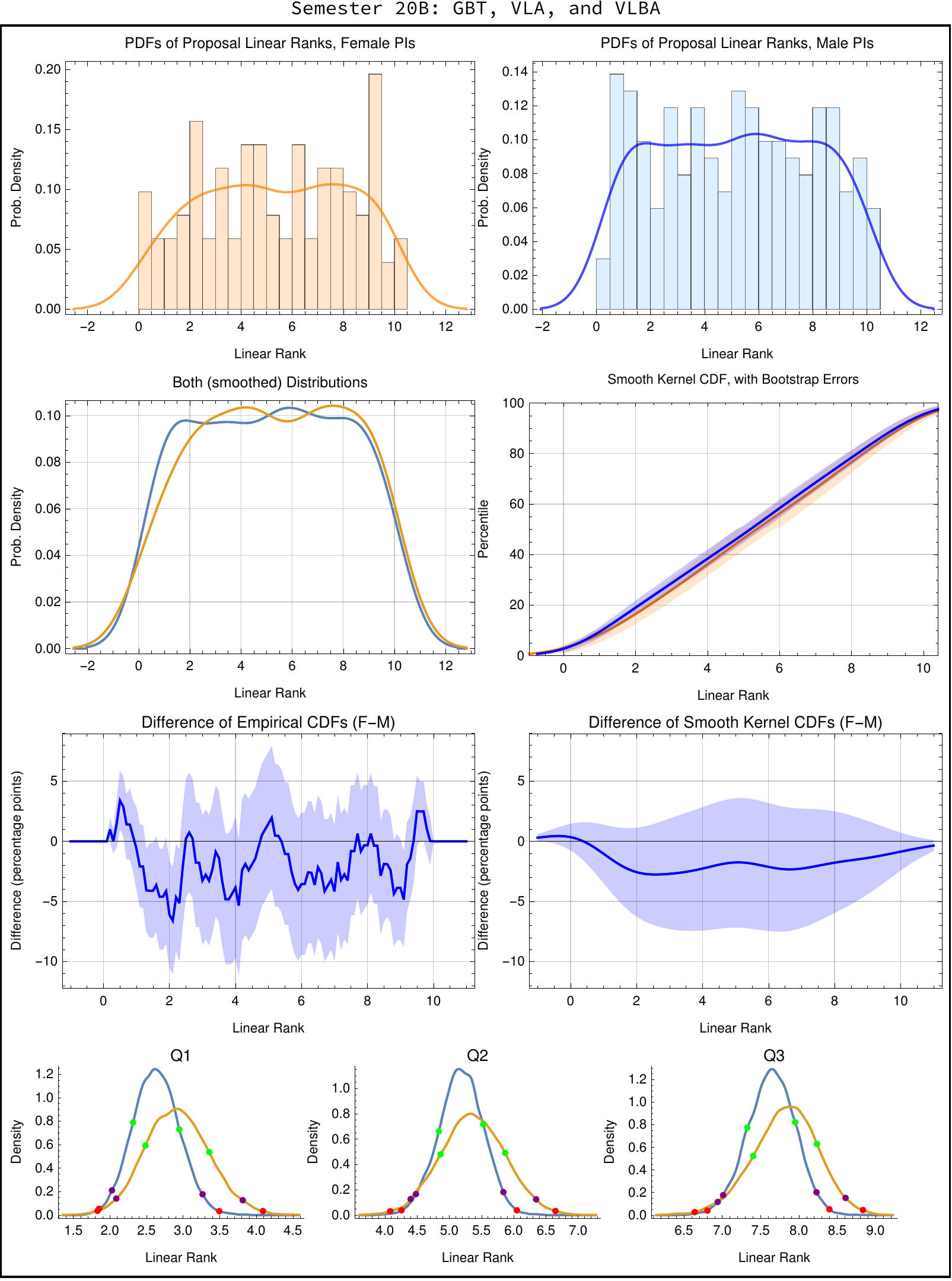}
  \caption{Statistics for semester 20B\@. See
    Figure~\ref{fig:stats_12a-17a} for details.}
\label{fig:stats_20b}
\end{figure}

\begin{figure}
  \centering
  \includegraphics[angle=0,scale=0.9]{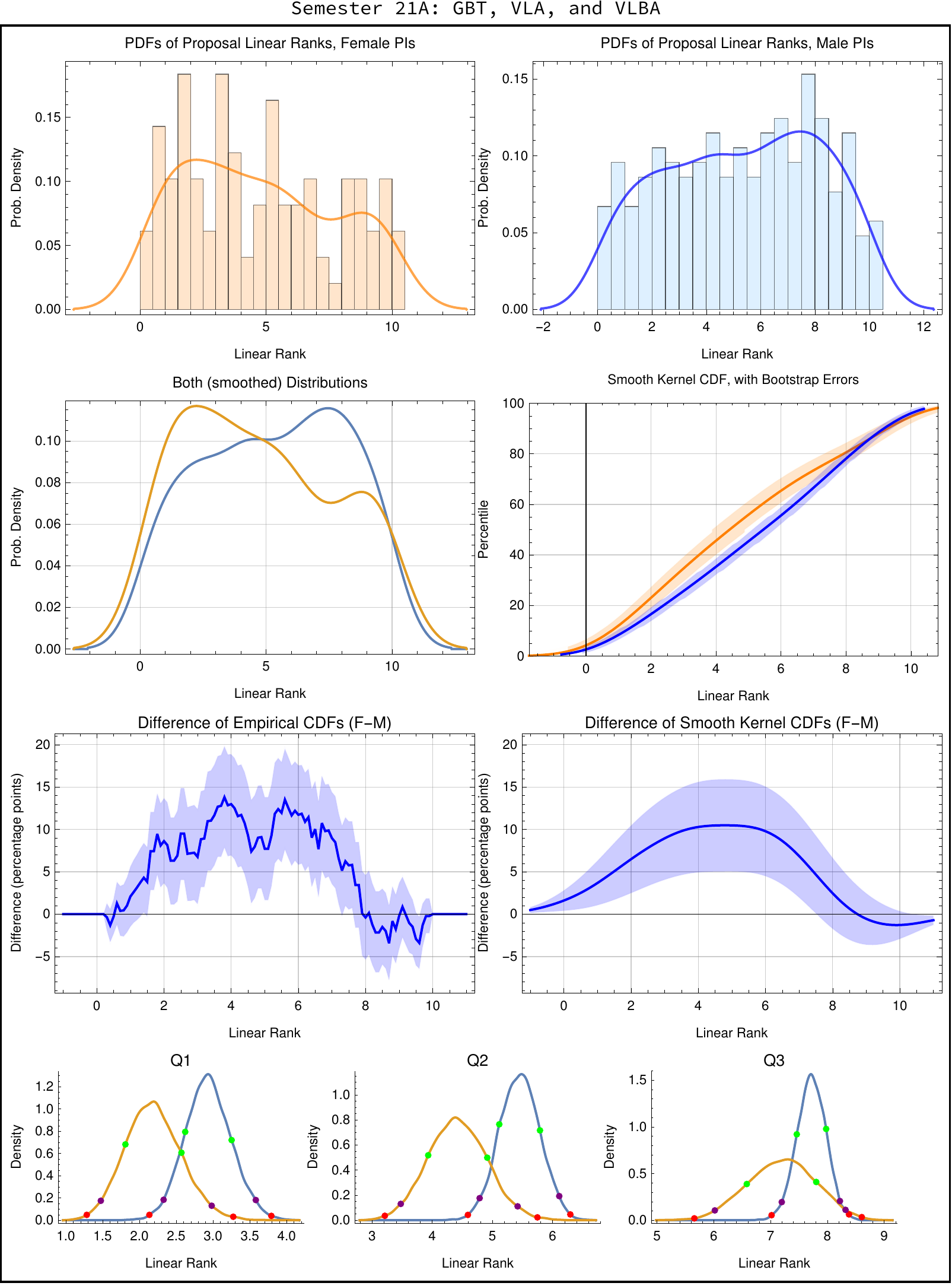}
  \caption{Statistics for semester 21A\@. See
    Figure~\ref{fig:stats_12a-17a} for details.}
\label{fig:stats_21a}
\end{figure}

\begin{figure}
  \centering
  \includegraphics[angle=0,scale=0.9]{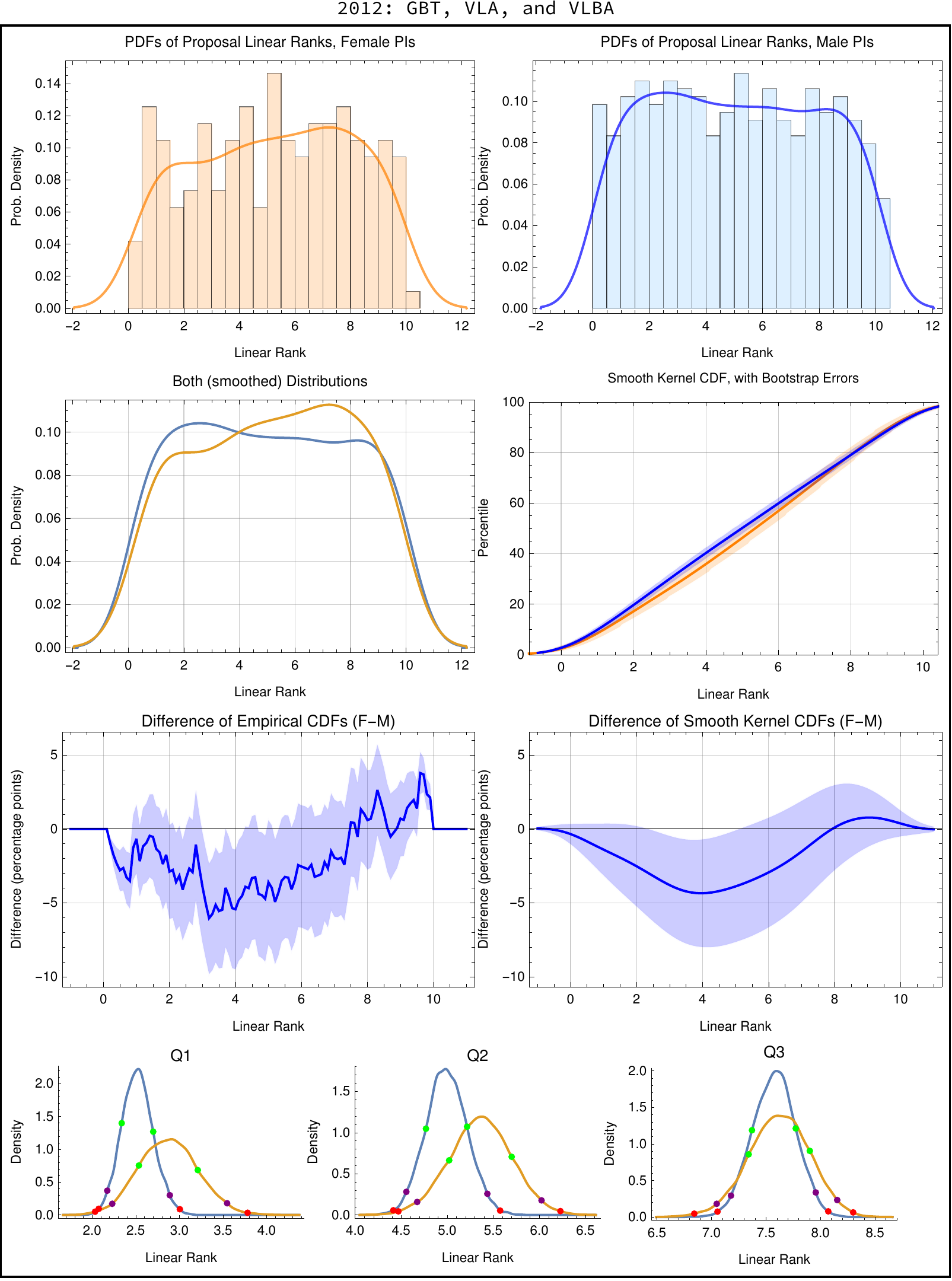}
  \caption{Statistics for year 2012. See
    Figure~\ref{fig:stats_12a-17a} for details.}
\label{fig:stats_2012}
\end{figure}

\begin{figure}
  \centering
  \includegraphics[angle=0,scale=0.9]{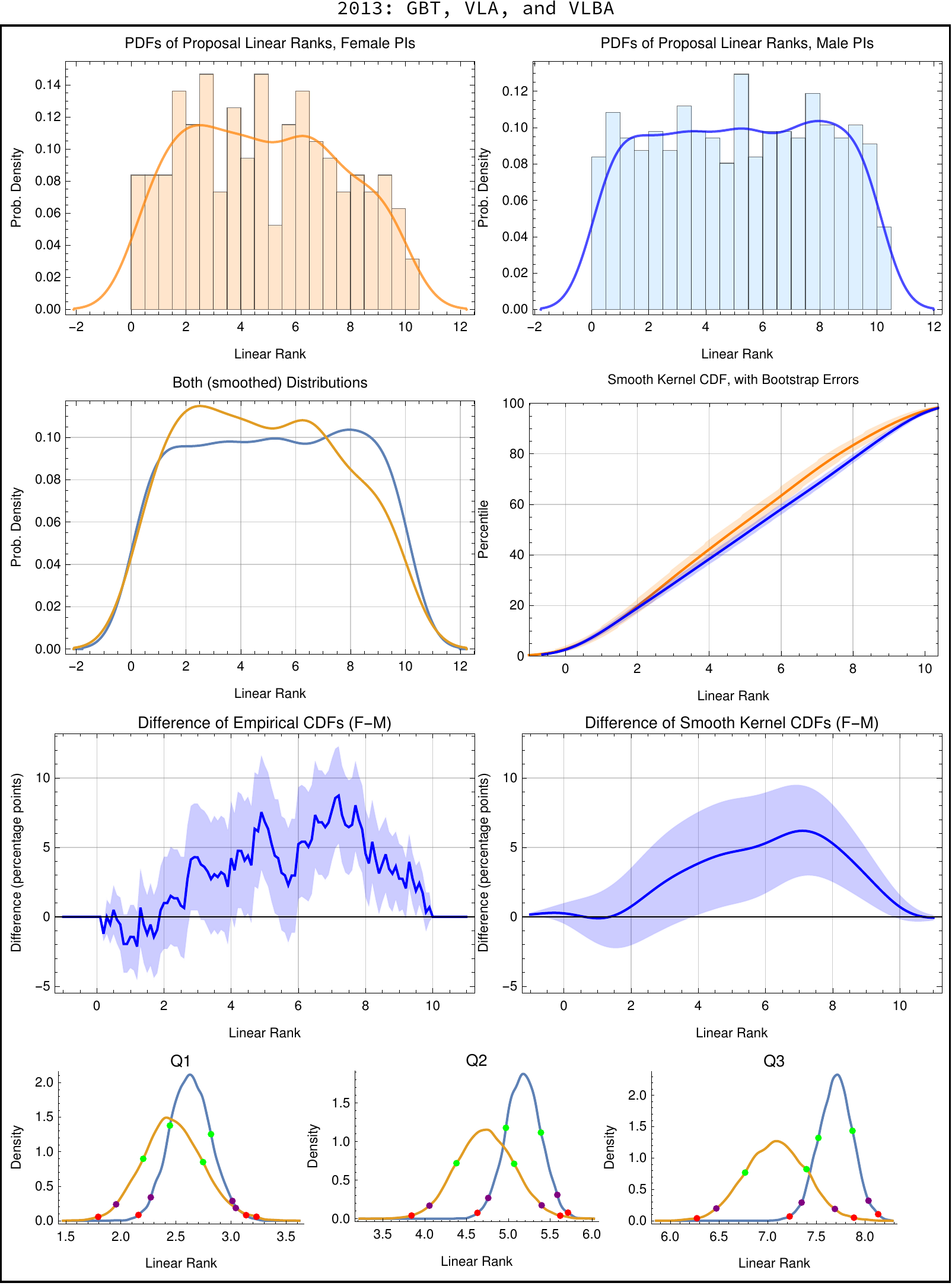}
  \caption{Statistics for year 2013. See
    Figure~\ref{fig:stats_12a-17a} for details.}
\label{fig:stats_2013}
\end{figure}

\begin{figure}
  \centering
  \includegraphics[angle=0,scale=0.9]{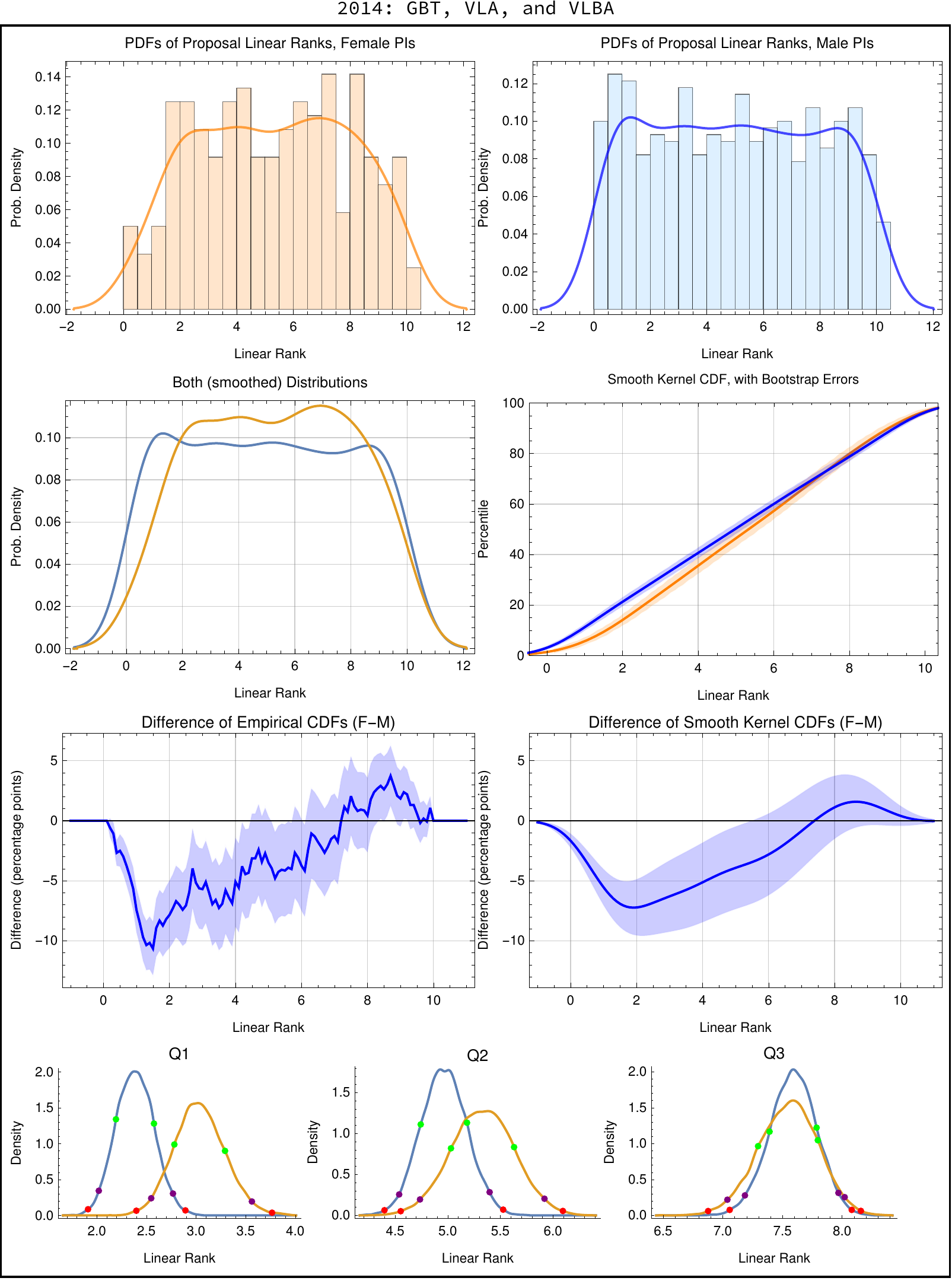}
  \caption{Statistics for year 2014. See
    Figure~\ref{fig:stats_12a-17a} for details.}
\label{fig:stats_2014}
\end{figure}

\begin{figure}
  \centering
  \includegraphics[angle=0,scale=0.9]{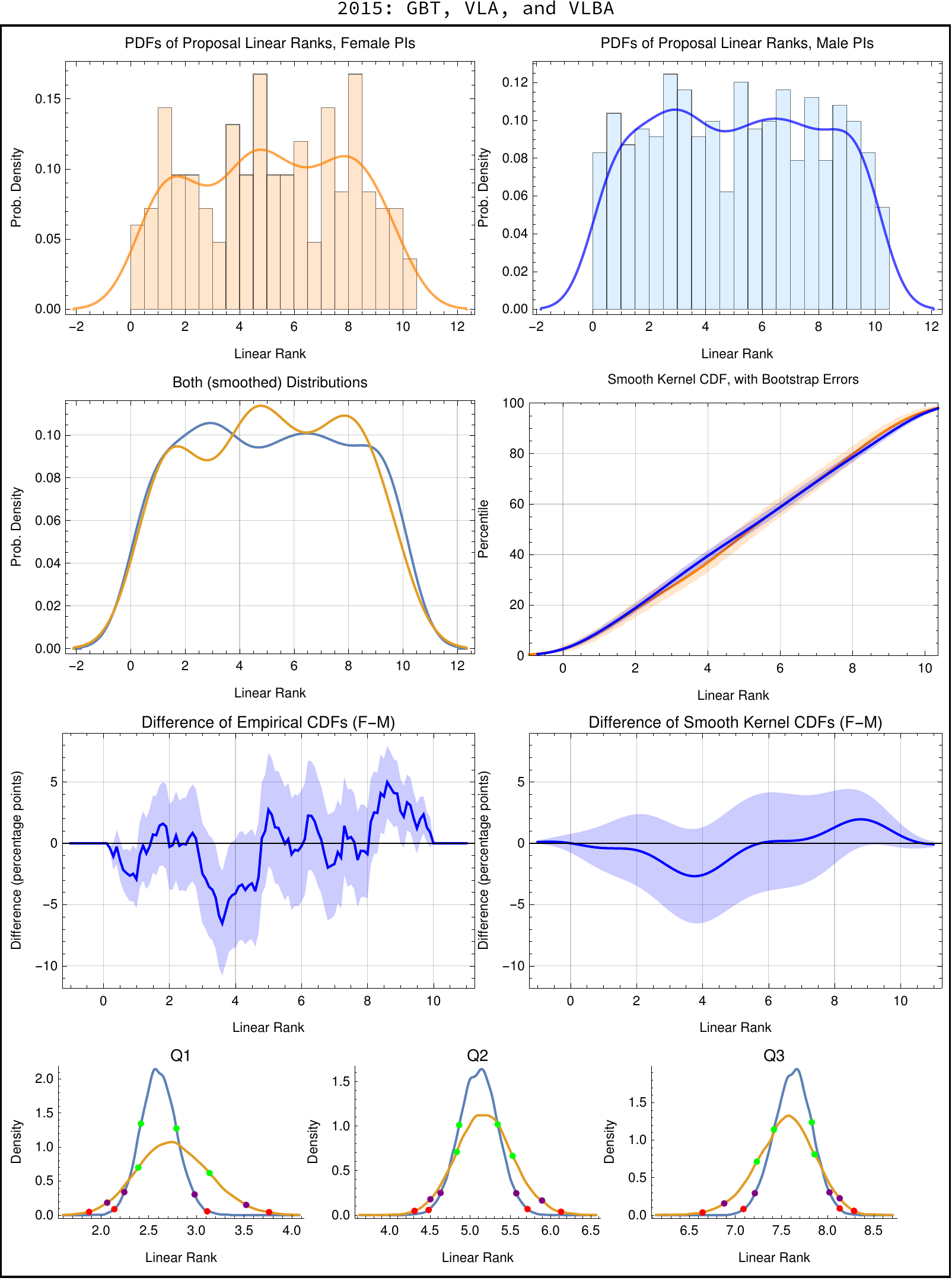}
  \caption{Statistics for year 2015. See
    Figure~\ref{fig:stats_12a-17a} for details.}
\label{fig:stats_2015}
\end{figure}

\begin{figure}
  \centering
  \includegraphics[angle=0,scale=0.9]{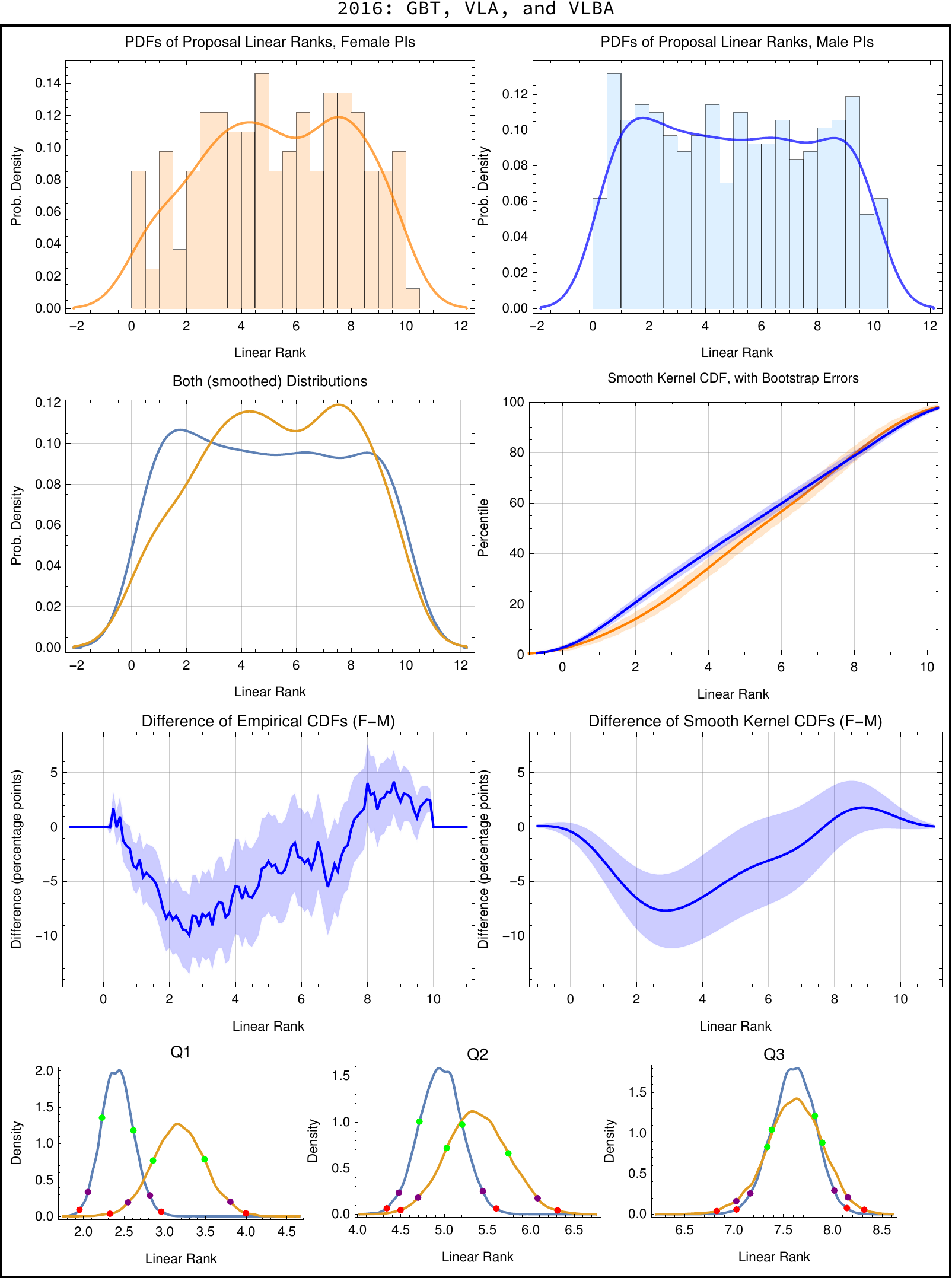}
  \caption{Statistics for year 2016. See
    Figure~\ref{fig:stats_12a-17a} for details.}
\label{fig:stats_2016}
\end{figure}

\begin{figure}
  \centering
  \includegraphics[angle=0,scale=0.9]{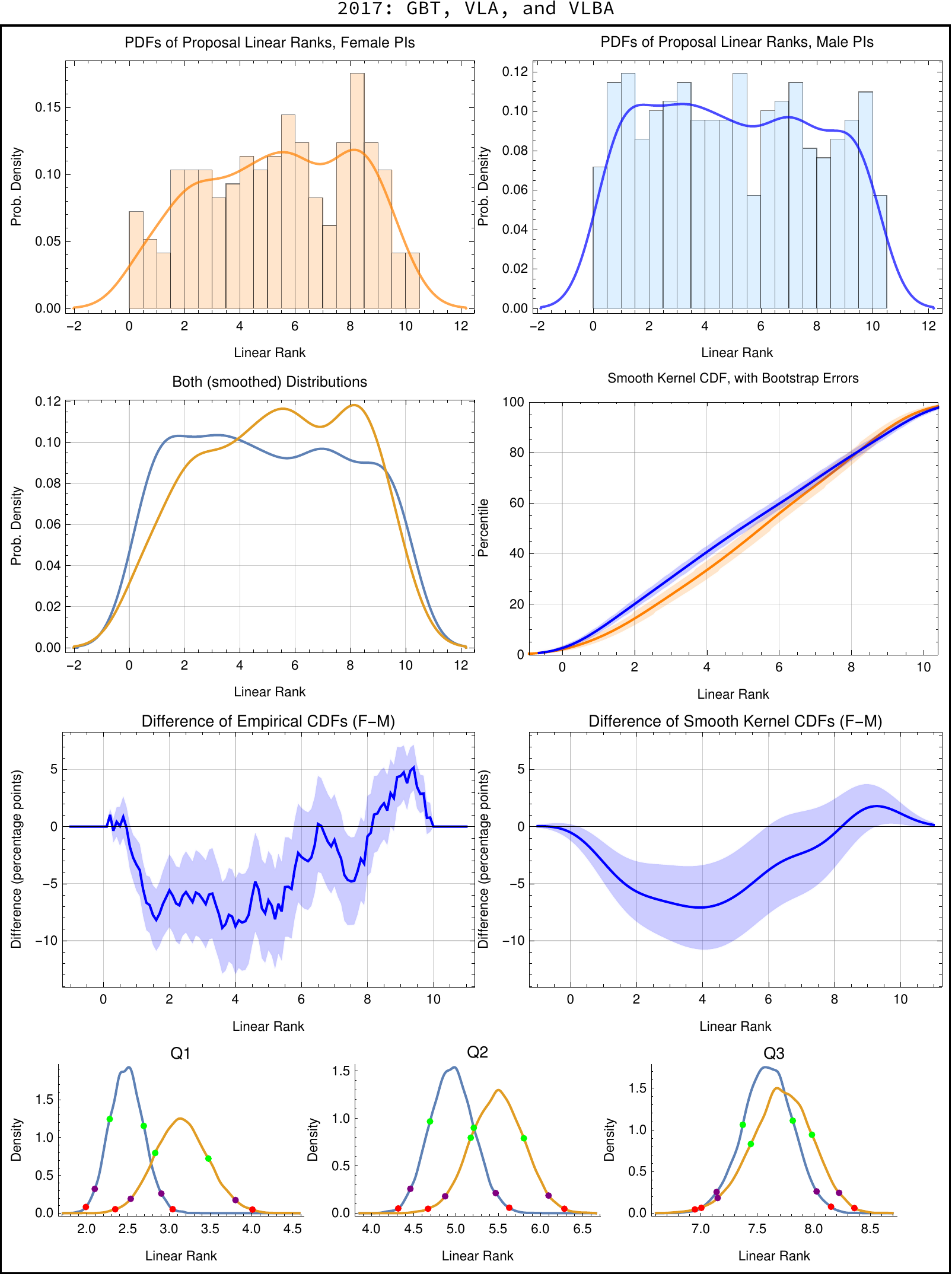}
  \caption{Statistics for year 2017. See
    Figure~\ref{fig:stats_12a-17a} for details.}
\label{fig:stats_2017}
\end{figure}

\begin{figure}
  \centering
  \includegraphics[angle=0,scale=0.9]{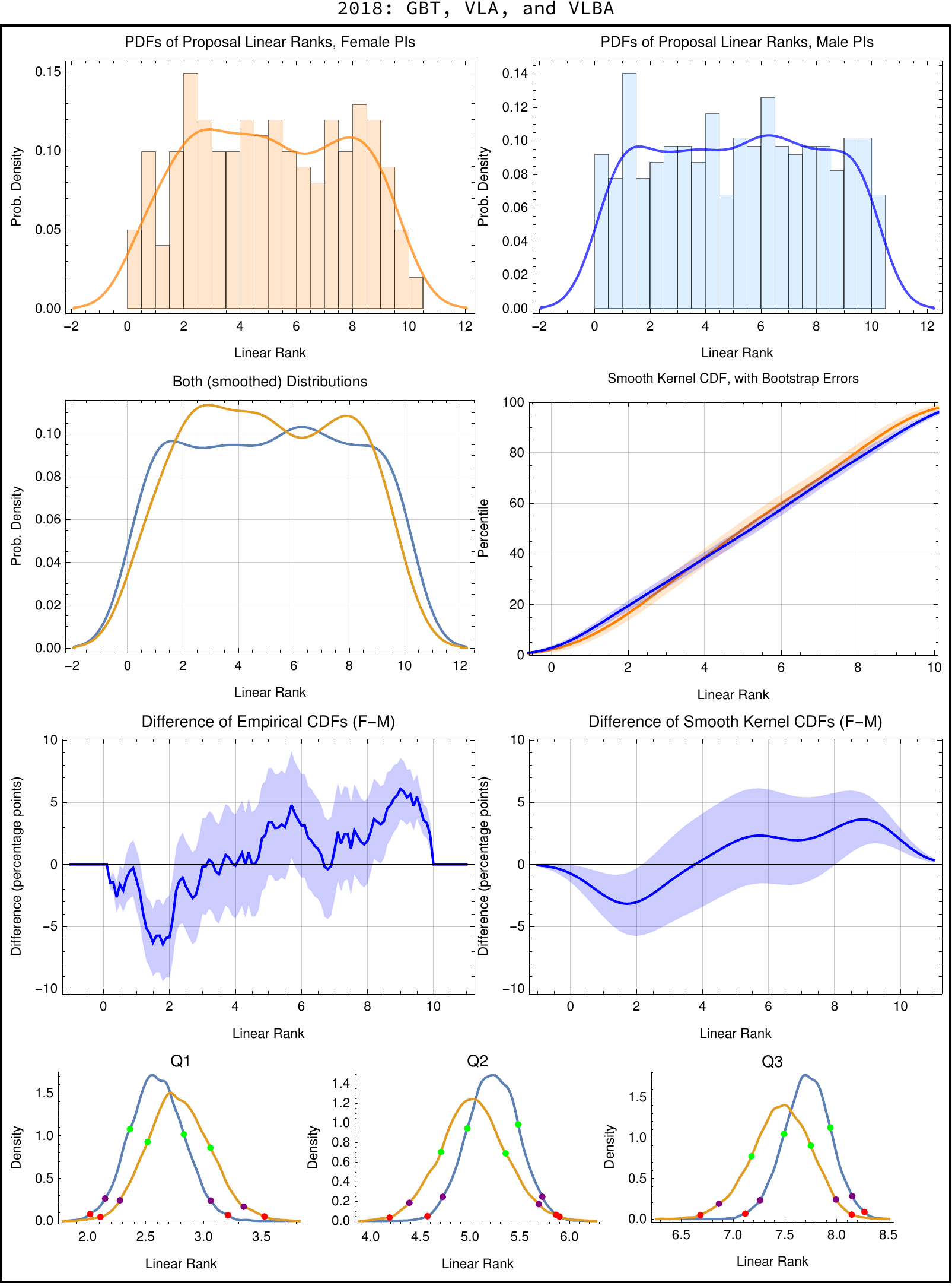}
  \caption{Statistics for year 2018. See
    Figure~\ref{fig:stats_12a-17a} for details.}
\label{fig:stats_2018}
\end{figure}

\begin{figure}
  \centering
  \includegraphics[angle=0,scale=0.9]{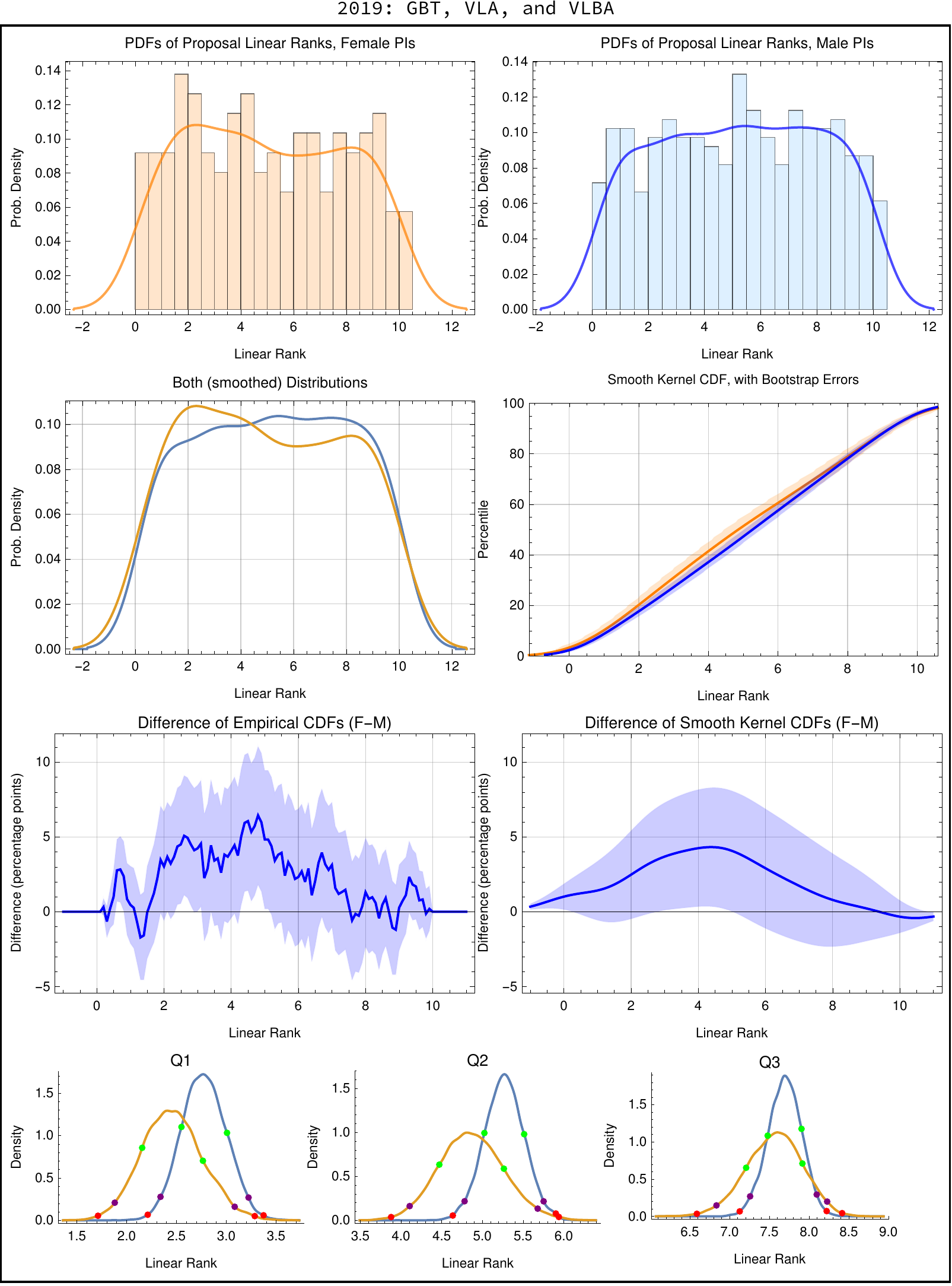}
  \caption{Statistics for year 2019. See
    Figure~\ref{fig:stats_12a-17a} for details.}
\label{fig:stats_2019}
\end{figure}

\begin{figure}
  \centering
  \includegraphics[angle=0,scale=0.9]{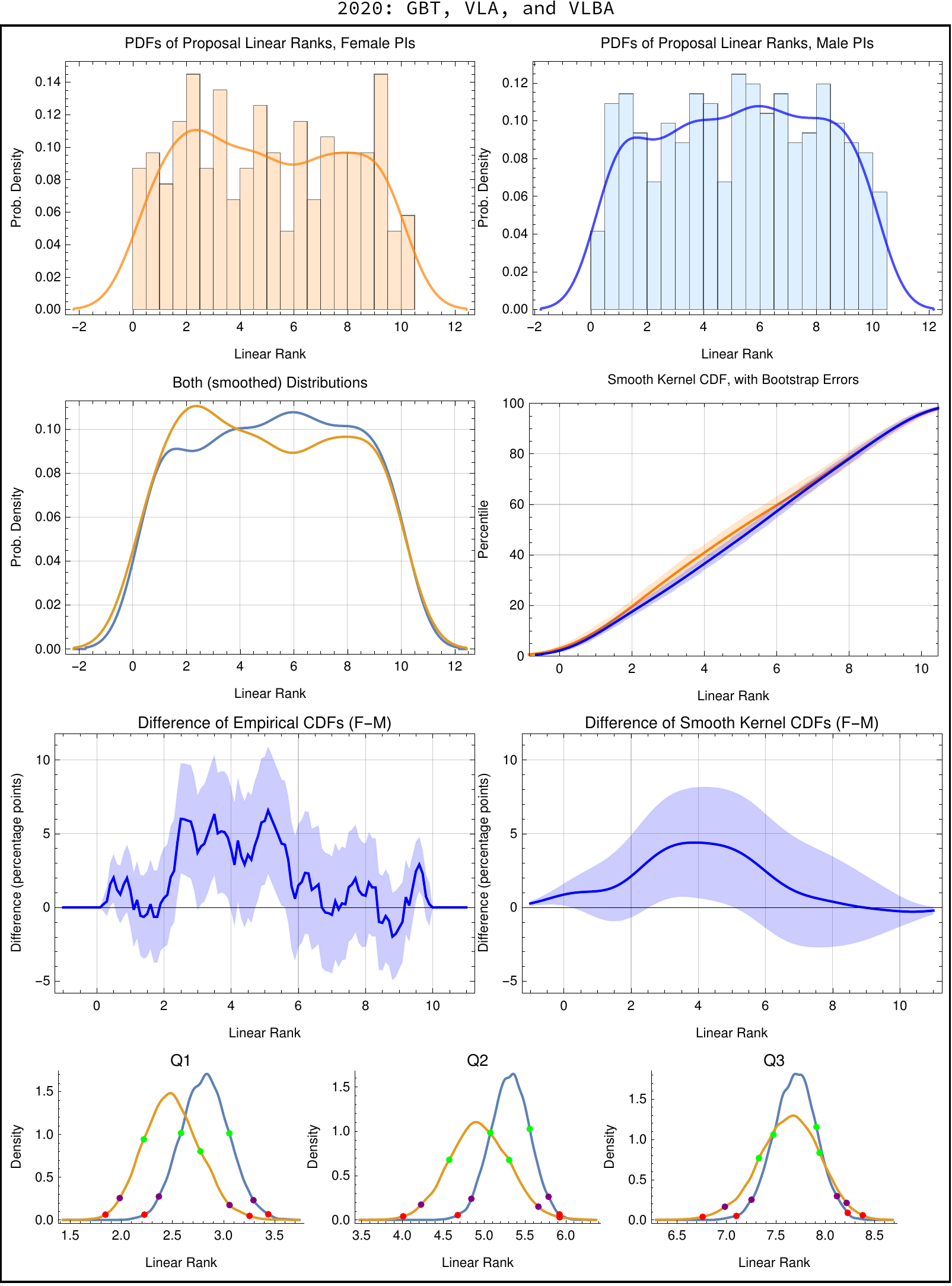}
  \caption{Statistics for year 2020. See
    Figure~\ref{fig:stats_12a-17a} for details.}
\label{fig:stats_2020}
\end{figure}

\begin{figure}
  \centering
  \includegraphics[angle=0,scale=0.9]{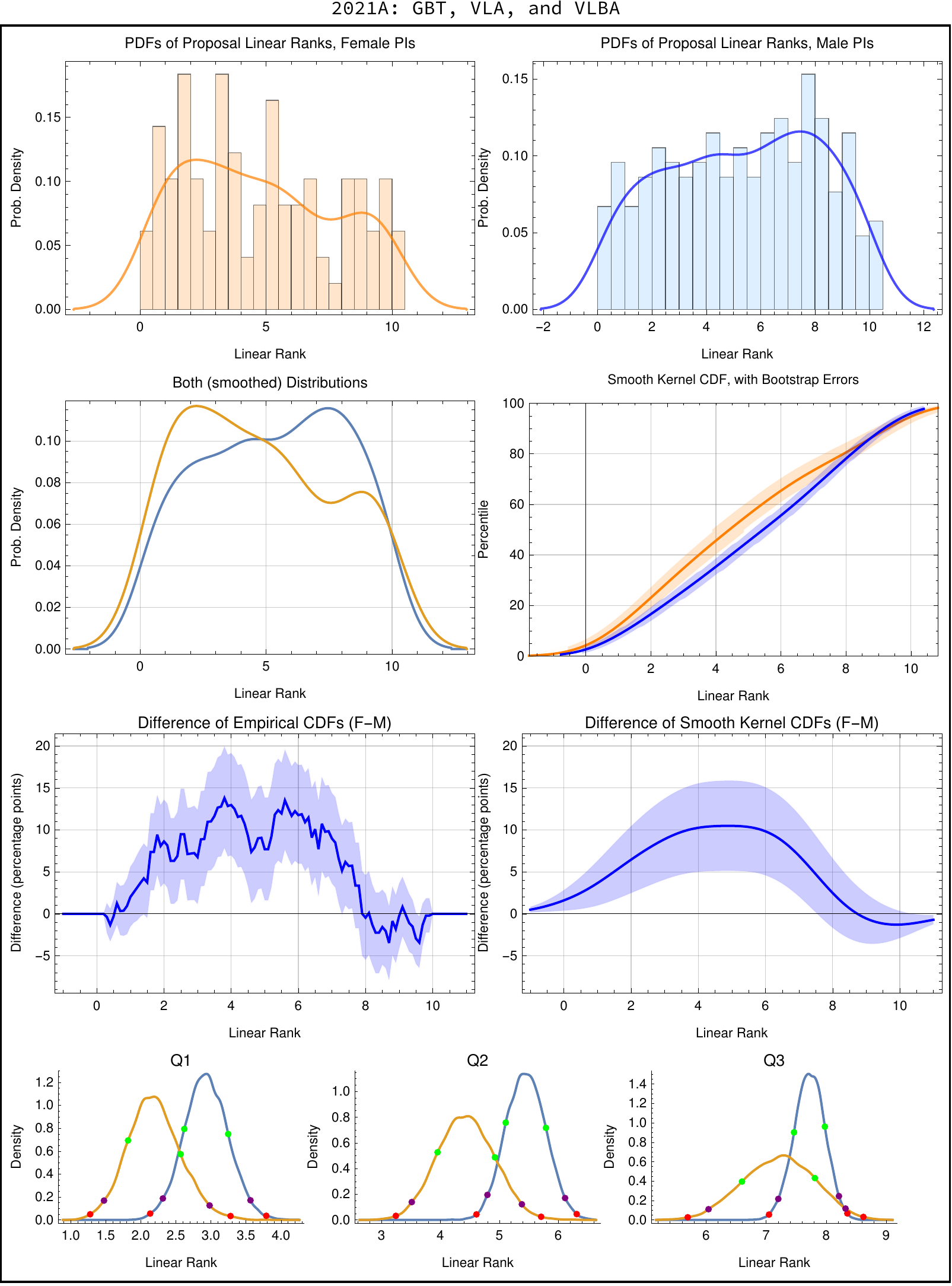}
  \caption{Statistics for year 2021 (only semester A is included). See
    Figure~\ref{fig:stats_12a-17a} for details.}
\label{fig:stats_2021}
\end{figure}

\begin{figure}
  \centering
  \includegraphics[angle=0,scale=0.9]{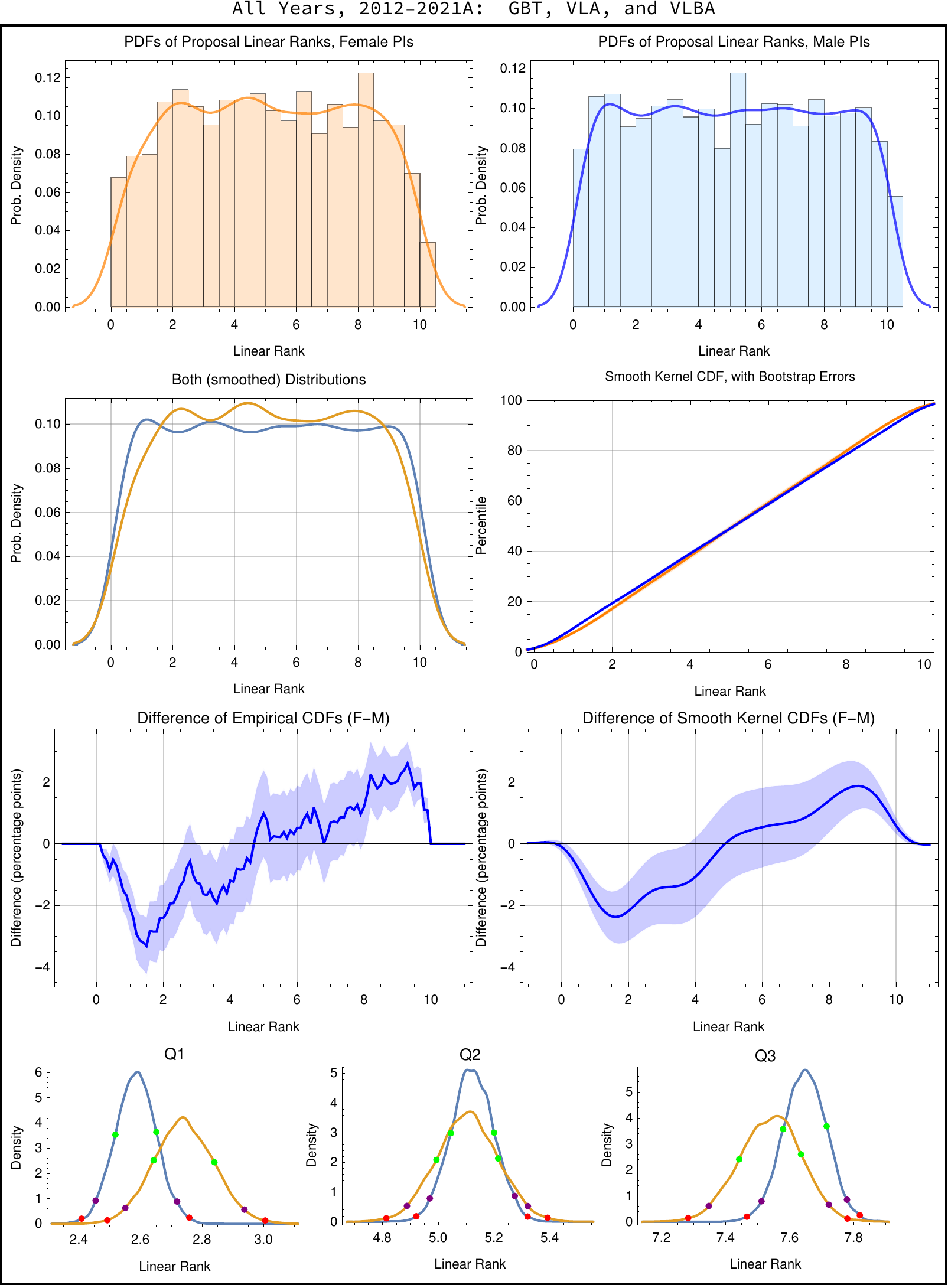}
  \caption{Statistics for years 2012--2021.  For the year 2021 only
    semester A is included.  See Figure~\ref{fig:stats_12a-17a} for
    details.}
\label{fig:stats_2012-2021}
\end{figure}

Here we summarize the results of our analysis for 30 data subsets that
include each semester
(Figures~\ref{fig:stats_12a}--\ref{fig:stats_21a}), each year
(Figures~\ref{fig:stats_2012}--\ref{fig:stats_2021}), and all years
(Figure~\ref{fig:stats_2012-2021}).  Plotted are the histograms, CDFs,
and quartile value distributions of the linear ranks for male and
female PIs.

\bibliography{gender}

\end{document}